\documentclass[12pt]{iopart}

\usepackage{graphicx}
\usepackage{color}

\begin{document}

\title[Universality class of the depinning transition ...]
{Universality class of the depinning transition in the two-dimensional Ising model with quenched disorder}

\author{X. P. Qin$^{1,2}$, B. Zheng$^{1}$ and N. J. Zhou$^{1}$}

\address{
$^1$ Department of Physics, Zhejiang University, Hangzhou 310027,
P.R. China\\
$^2$ School of Science, Zhejiang University of Science and
Technology, Hangzhou 310023, P.R. China}
 \ead{\mailto{zheng@zimp.zju.edu.cn}}

\begin{abstract}
With Monte Carlo methods, we investigate the universality class of
the depinning transition in the two-dimensional Ising model with
quenched random fields. Based on the short-time dynamic approach, we accurately
determine the depinning transition field and both static and dynamic
critical exponents. The critical exponents
vary significantly with the form and strength of the random
fields, but exhibit independence on the updating schemes of the Monte Carlo algorithm.
From the roughness exponents $\zeta, \zeta_{loc}$ and $\zeta_s$,
one may judge that the depinning
transition of the random-field Ising model belongs to the new
dynamic universality class with $\zeta \neq \zeta_{loc}\neq \zeta_s$
and $\zeta_{loc} \neq 1$. The crossover from the
second-order phase transition to the first-order one is observed
for the uniform distribution of the random fields, but it is not present
for the Gaussian distribution.
\end{abstract}


\maketitle

\section{Introduction}

In recent years many activities have been devoted to the study of dynamic
processes far from equilibrium. An example is the domain-wall dynamics in
disordered media, which is important from both experimental and theoretical
perspectives. A crucial feature of the driven domain interface in
disordered media is the {\it depinning} phase transition at zero
temperature. For the domain-wall motion in ultrathin magnetic films,
the Edwards-Wilkinson equation with quenched disorder (QEW) is a typical
theoretical approach \cite{nat01,gla03,ros03,due05,kol06,kol06a,kle07a,bus08}.
It is usually believed that different variants of the model
belong to a same universality class.

The QEW equation is a phenomenological model, without detailed microscopic
structures and interactions of the materials. The
domain wall in a two-dimensional system is effectively
described by a single-valued elastic string.
An alternative, possibly more realistic approach, may be to construct
lattice models at the microscopic level. The two-dimensional random-field Ising model with a driving field
(DRFIM) is the simplest example \cite{now98,rot99,rot01,zho09,zho10}.
In the DRFIM model, overhangs and islands may be created during the dynamic evolution,
and the domain wall is not single-valued and one dimensional.
The DRFIM model is closer to experiments \cite{met07,lem98,jos98} and does not suffer from the theoretical self-inconsistence
as in the QEW equation \cite{ros01}.
Very recently it has been demonstrated that the DRFIM model may not belong to the universality
class of the QEW equation \cite{zho09,zho10}.
To distinguish these two models, one needs accurate measurements of the critical exponents.
In this respect, the short-time dynamic approach has been proven to be efficient \cite{kol06,zhe98,luo98,rod07,kol09,zho09,zho10}.

The purpose of this paper is to clarify the universality class of
the DRFIM model. In Refs.~\cite{zho09,zho10}, Monte Carlo
simulations are performed at a fixed strength of the random fields with a uniform distribution,
i.e., $\Delta=1.5$. In this paper, we systematically investigate the possible
dependence of the critical exponents on the strength and form of the random
fields, and the updating schemes of the Monte Carlo algorithm, including the crossover from the second-order transition to
the first-order one. On the other hand, we measure not only the
global and local roughness exponents $\zeta$ and $\zeta_{loc}$,  but
also the {\it spectral} roughness exponent $\zeta_s$, to identify
the dynamic universality class of the domain interface. Following
Ref. \cite{ram00}, there are four different universality classes in
the interface growth, namely,
\begin{equation}
   \begin{array}{lll}
      \mbox{if} ~~ \zeta_s  < 1, \ \zeta_{loc}=\zeta_s ~
      \left\{
       \begin{array}{lll}
          \zeta_s=\zeta  ~ \quad ~ \mbox{Family-Vicsek}  \\
          \zeta_s \neq \zeta  ~ \quad ~ \mbox{intrinsic,}
       \end{array}\right.  \\
     \mbox{if} ~~ \zeta_s > 1, \  \zeta_{loc}=1 ~~
      \left\{
       \begin{array}{lll}
          \zeta_s=\zeta  ~ \quad ~ \mbox{super-rough}  \\
          \zeta_s \neq \zeta  ~ \quad ~ \mbox{faceted.}
       \end{array}\right.
   \end{array}
   \label{equ00}
\end{equation}
In Ref.~\cite{che10}, however, it suggests that there is a new
universality class of {\it anomalous roughening}, with $\zeta \neq
\zeta_{loc} \neq \zeta_s$ and $\zeta_{loc} \neq 1$.
We verify that the DRFIM model belongs to such a
new dynamic universality class.
In section 2, the model and scaling analysis are described. In section 3,
numerical results are presented. Section 4 includes the conclusions.

\section{Model and scaling analysis}

\subsection{Model}
The two-dimensional (2D) random-field Ising model with a driving
field is defined by the Hamiltonian
\begin{eqnarray}
\mathcal{H} &=& - J \sum_{<ij>}S_i S_j  - \sum_i
(h_i+H) S_i . \label{equ10}
\end{eqnarray}
where $S_i=\pm 1$ is an Ising spin at site $i$ of a rectangle
lattice $L_x\times L_y$, $<ij>$ denotes a summation over nearest
neighbors and $H$ is a homogeneous driving field.
The random-field $h_i$ may be distributed in different forms.
A typical example is a uniform distribution within an interval
$[-\Delta,\Delta]$, following Refs.~\cite{now98,zho09,zho10}. We
take $\Delta$ from $0.8J$ to $2.3J$ and set the coupling constant
$J=1$. Another example is a Gaussian distribution with mean zero and standard deviation
$\sigma$. Different form and strength of the random fields may lead to different
kinds of phase transitions \cite{ji91,koi10}. At large $\Delta$ or $\sigma$, the
depinning transition is of second-order for both the uniform
and Gaussian distributions. As $\Delta$ decreases, the
depinning transition crosses over to a first-order one for the uniform
distribution which is bounded. However, such a crossover is not present for the Gaussian
distribution which is unbounded.

We first concentrate on the uniform distribution of the random fields and the random-single-spin-flip Monte Carlo algorithm,
and then extend the calculations to the Gaussian distribution and to different updating schemes.
Simulations are performed
at zero temperature with lattice sizes $L_y=256$, $512$, $1024$, and
$2048$ up to $t_{max}=1280, 2560$ or $5120$ for different random
fields. Total samples for average are about 20000. Main results are
presented with $L_y=1024$ and simulations of different $L_y$ confirm
that finite-size effects are already negligibly small. $L_x$ is
taken to be sufficiently large so that the boundary is not reached.
Statistical errors are estimated by dividing the total samples in
a few subgroups. If the fluctuation in the time direction is
comparable with or larger than the statistical error, it will be
taken into account.

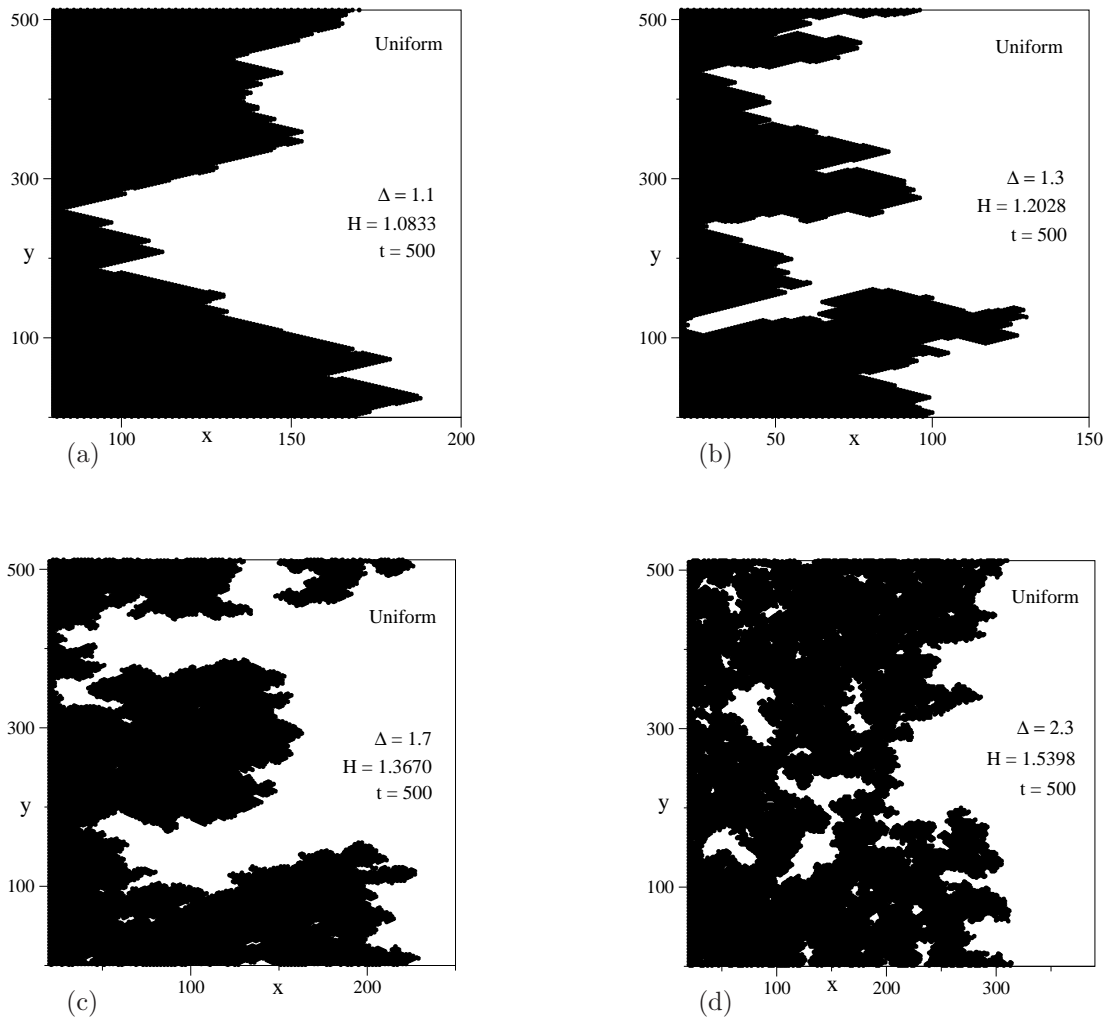
\begin{figure}[ht]
\setlength{\unitlength}{1.cm}
\begin{picture}(6,16)(0,-9)
\put(0,-8.6){\includegraphics[width=6.cm]{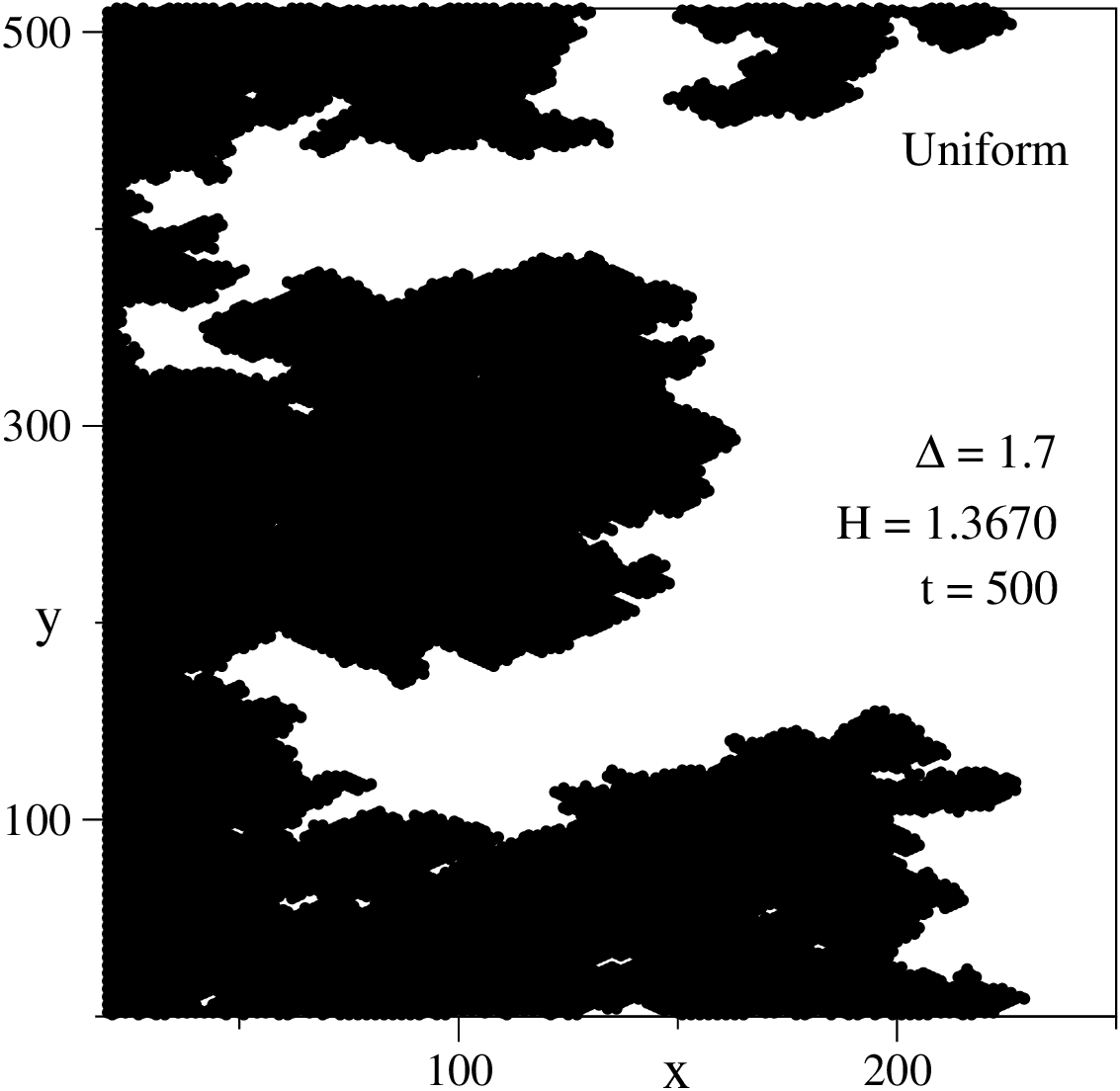}}
\put(8.5,-8.6){\includegraphics[width=6.cm]{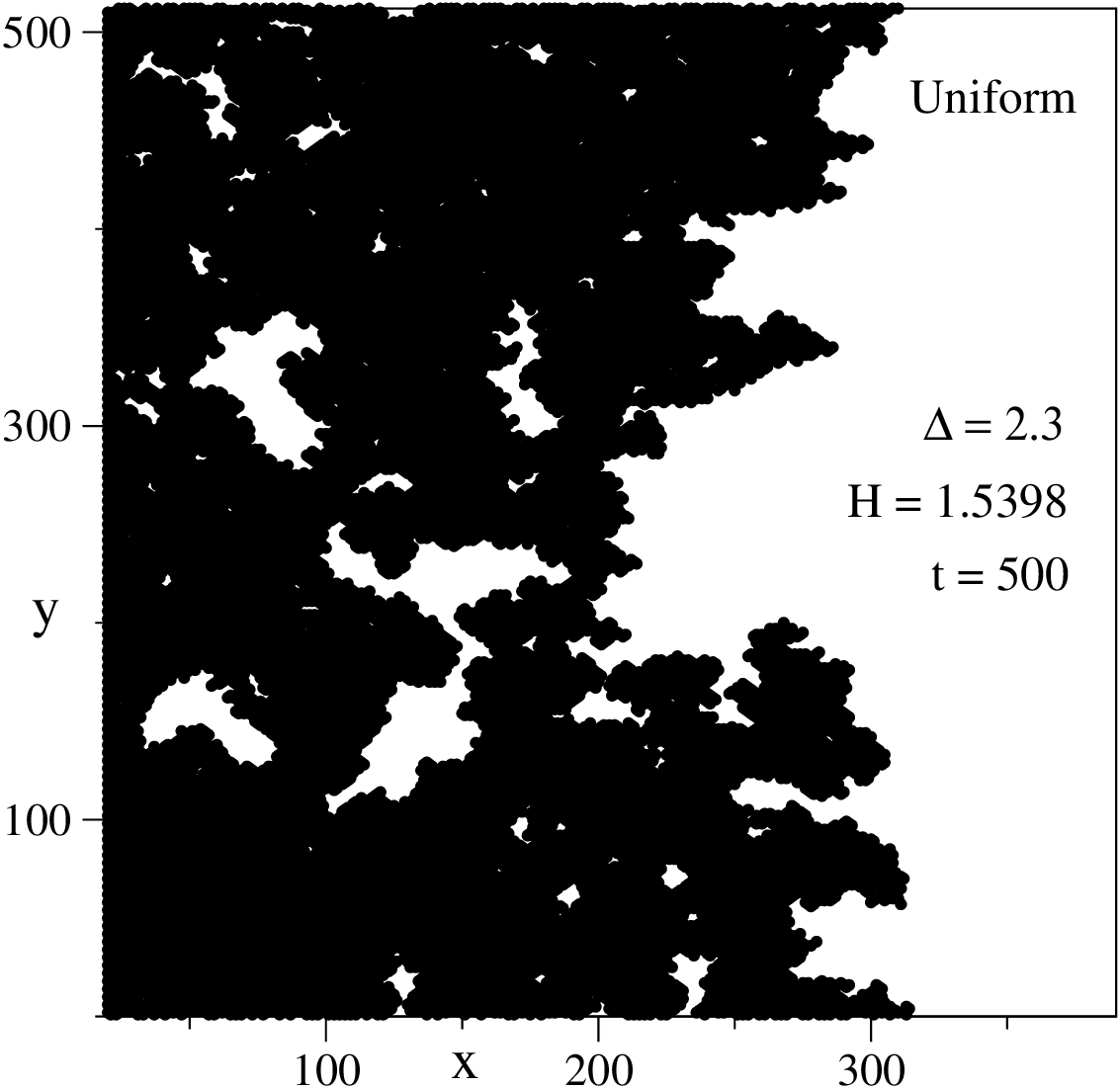}}
\put(0.8,-8.8){\footnotesize(c)} \put(9.2,-8.8){\footnotesize(d)}
\put(0.05,-1.3){\includegraphics[width=6.2cm]{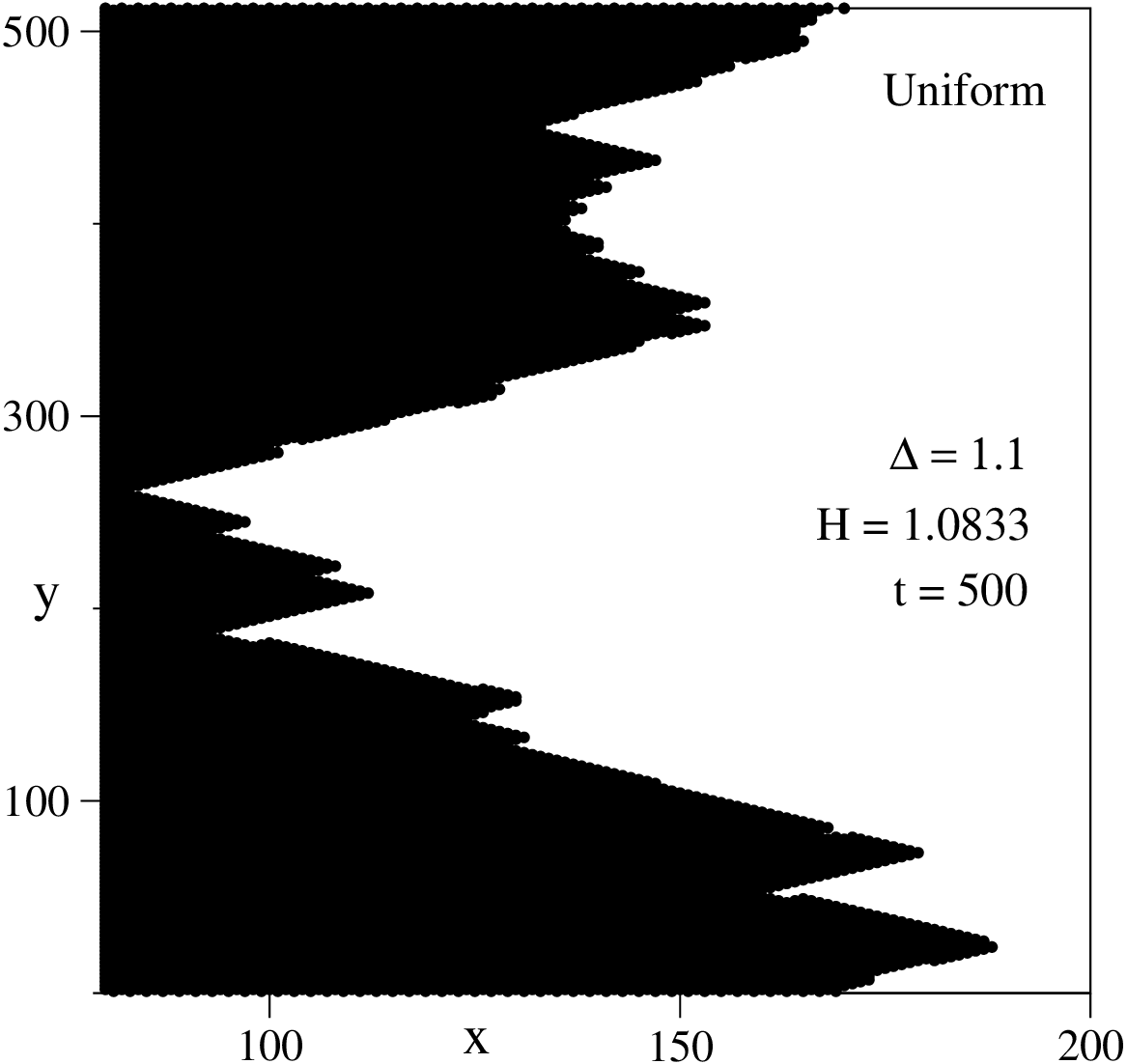}}
\put(8.4,-1.3){\includegraphics[width=6.2cm]{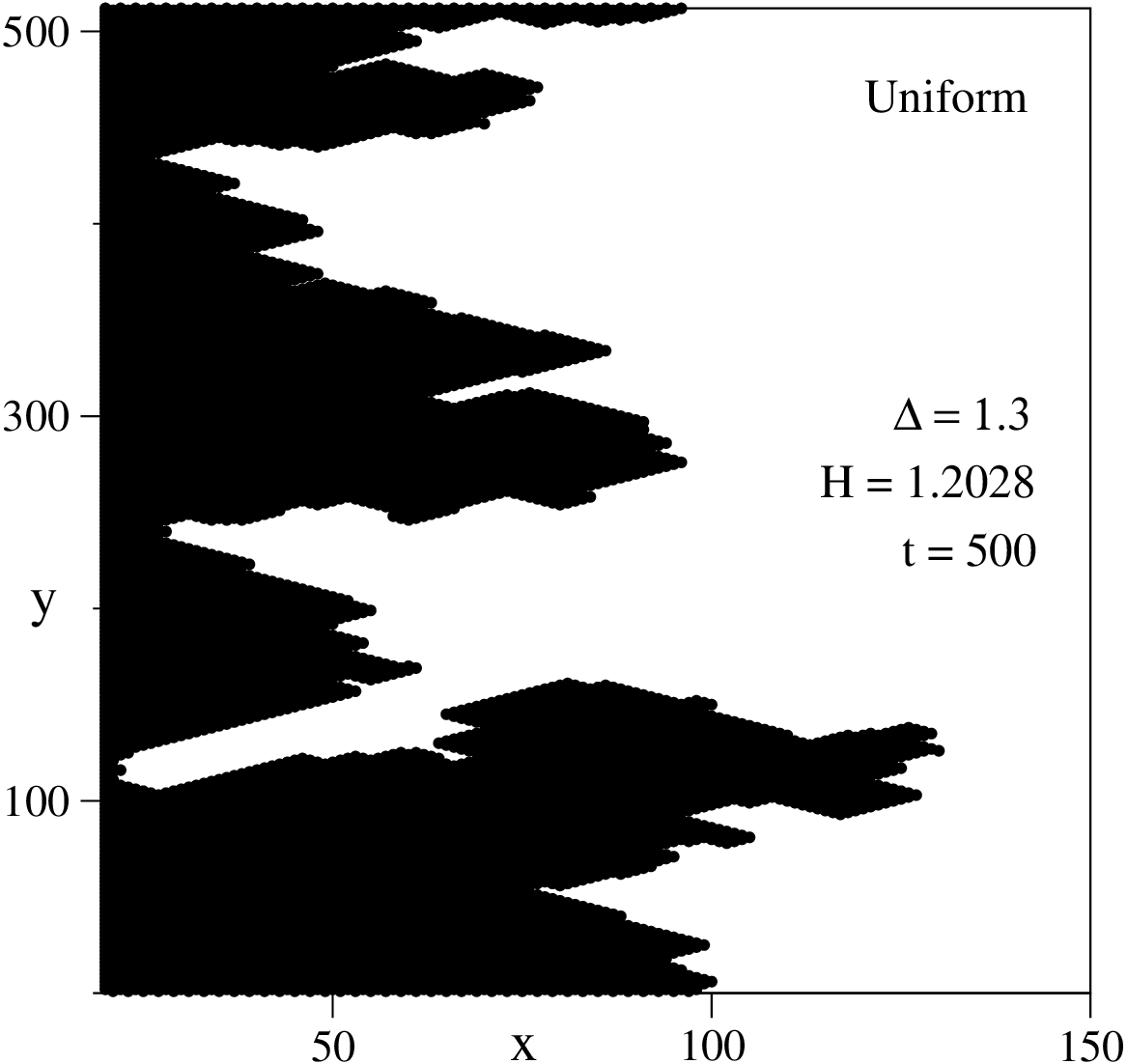}}
\put(0.8,-1.5){\footnotesize(a)} \put(9.2,-1.5){\footnotesize(b)}
\end{picture}
\caption{Snapshots of the spin configuration for the uniform distribution of the random fields. The black denotes
the spin $S_i=1$ and the white denotes the spin $S_i=-1$. Simulations are performed with the random-single-spin-flip Monte Carlo algorithm.}
\label{evolution}
\end{figure}

The initial state is a {\it semi-ordered} state with a perfect domain
wall in the $y$ direction. The periodic boundary condition is used
in $y$ direction, while the spins at the boundary in $x$ direction
are fixed. To eliminate the pinning effect irrelevant for disorder,
we rotate the lattice such that the initial domain wall orients in
the $(11)$ direction of the lattice ~\cite{now98}. After preparing
the initial state, the {\it random-single-spin-flip} Monte Carlo algorithm is performed.
We $randomly$ select {\it one spin}, and flip it if the
total energy decreases after flipping, and if at least
one of the nearest neighbors is already flipped, i.e., only spins at the
interface are allowed to flip \cite{koi10,dro98}. A Monte Carlo time step
is defined as $L_x \times L_y$ single spin flips. As time
evolves, the domain wall propagates and roughens. In
Fig.~\ref{evolution}, the snapshots of the dynamic evolution of the
spin configuration are displayed for different random fields with the
uniform distribution. As $\Delta$ increases, overhangs first appear
then islands.

Due to the existence of the overhangs and islands, there may be
different ways to define the domain interface. Here, we adopt the
definition with the magnetization. Denoting a spin at site $(x,y)$
by $S_{xy}(t)$, we introduce a $line$ magnetization and height
function of the domain interface \cite{zho09,zho10},
\begin{equation}
m(y,t) = \frac{1}{L_x} \left[ \sum_{x=1}^{L_x} S_{xy}(t) \right],
\label{equ20}
\end{equation}
\begin{equation}
h(y,t) = \frac{L_x}{2}[ m(y,t) + 1] . \label{equ30}
\end{equation}
We then calculate the average velocity
and roughness function of the domain interface,
\begin{equation}
v(t) = \frac{d\langle h(y,t)\rangle}{dt}, \label{equ40}
\end{equation}
\begin{equation}
\omega^{2}(t) = \left \langle h(y,t)^2 \right \rangle - \langle
h(y,t) \rangle^2. \label{equ50}
\end{equation}
Here $\langle \cdots \rangle$ represents both the statistical
average and the average in $y$ direction.

A more informative quantity is the height correlation function,
\begin{equation}
C(r, t) = \left\langle[h(y + r, t) - h(y, t)]^2 \right \rangle.
\label{equ60}
\end{equation}
It describes both the spatial correlation of the height function in
$y$ direction and the growth of the domain interface in $x$
direction. Further, we consider the Fourier transform of the height
function \cite{ram00,che10},
\begin{equation}
h(k,t) = \frac{1}{\sqrt{L_y}} \sum_{y=1}^{L_y} \left[ h(y,t)-
\langle h(y,t) \rangle \right]exp(iky). \label{equ70}
\end{equation}
The structure factor is then defined,
\begin{equation}
S(k,t) = \langle h(k,t)h(-k,t) \rangle . \label{equ80}
\end{equation}

To obtain the dynamic exponent $z$ {\it independently}, we introduce
an observable
\begin{equation}
F(t) = [M^{(2)}(t) - M(t)^2]/\omega^2(t). \label{equ90}
\end{equation}
Here $M(t)$ is the global magnetization and $M^{(2)}(t)$ is its
second moment. $F(t)$ is nothing but the ratio of the
planar susceptibility and line susceptibility.

\subsection{Scaling analysis}

For the uniform distribution with $\Delta>1$ and Gaussian distribution of the random fields, the depinning transition is
a second-order phase transition. The order parameter, i.e., the
average velocity $v(t)$ should obey the dynamic scaling theory
supported by the renormalization-group calculations
\cite{jan89,zhe98,luo98}. In the critical regime, there are two
spatial length scales in the dynamic relaxation, i.e., the
nonequilibrium spatial correlation $\xi(t)$ and the finite lattice
size $L_y$, scaling arguments lead to a dynamic scaling form
\cite{jan89,zhe98,luo98},
\begin{equation}
v(t, \tau, L_y) = b^{-\beta/\nu} G(b^{-z}t, b^{1/\nu}\tau,
b^{-1}L_y). \label{equ100}
\end{equation}
Here, $b$ is a rescaling factor, $\beta$ and $\nu$ are
the static exponents, $z$ is the dynamic exponent, and
$\tau=(H-H_c)/H_c$. Taking $b \sim t^{1/z}$, the dynamic scaling
form is rewritten as
\begin{equation}
v(t, \tau, L_y) = t^{-\beta/\nu z} G(1, t^{1/\nu z}\tau,
t^{-1/z}L_y). \label{equ110}
\end{equation}
In the short-time regime, i.e., the regime with $\xi(t)\sim t^{1/z}
\ll L_y$, the finite-size effect is negligibly small,
\begin{equation}
v(t, \tau) = t^{-\beta / \nu z}G(t^{1/\nu z}\tau). \label{equ120}
\end{equation}
At the depinning transition point $\tau = 0$, a
power-law behavior is obtained,
\begin{equation}
v(t) \sim t^{-\beta / \nu z}. \label{equ130}
\end{equation}
With Eq.~(\ref{equ120}), one may locate the transition field $H_c$ by
searching for the best power-law behavior of $v(t, \tau)$
\cite{zhe98,luo98}. To determine $\nu$, one simply derives from
Eq.~(\ref{equ120})
\begin{equation}
\partial _\tau \ln v(t,\tau)|_{\tau=0} \sim t^{1 / \nu z}. \label{equ140}
\end{equation}
In general, the roughness
function $\omega^2(t)$ and height correlation function $C(r,t)$
may not obey a perfect power-law behavior in early times. Due to the
random updating scheme in numerical simulations, the domain
interface and its velocity also roughen {\it even without disorder}
($\Delta=0$). This leads to rather strong correlations to scaling.
To capture the dynamic effects of the disorder, we introduce the
pure roughness function
\begin{equation}
D\omega^2(t) = \omega^2(t) - \omega^2_b(t), \label{equ150}
\end{equation}
and height correlation function
\begin{equation}
DC(r,t) = C(r,t)- C_b(r,t), \label{equ160}
\end{equation}
where $\omega^2_b(t)$ and $C_b(r,t)$ are the roughness function and
height correlation function for $\Delta =0$. For a
large lattice $L_y$ and at the transition point $H=H_c$,
$D\omega^2(t)$ and $DC(r,t)$ should obey the standard power-law
scaling behaviors \cite{jos96,zho08,bak08,zho09},
\begin{equation}
D\omega^2(t) \sim t^{ 2 \zeta / z}, \label{equ170}
\end{equation}
and
\begin{equation}
   DC(r,t) \sim  \left\{
   \begin{array}{lll}
     t^{2(\zeta-\zeta_{loc})/z}\ r^{2\zeta_{loc}}   & \quad &  \mbox{if ~ $r \ll \xi(t) \ll L_y$} \\
     t^{2\zeta /z}  & \quad &  \mbox{if ~ $0 \ll \xi(t) \ll r$}
   \end{array}\right. .
   \label{equ180}
\end{equation}
Here $\xi(t) \sim t^{1/z}$, $\zeta$ is the global roughness
exponent, and $\zeta_{loc}$ is the local one.

The structure factor should follow a scaling form\cite{ram00,che10}
\begin{equation}
S(k,t) = k^{-(2 \zeta +1)} f_s(kt^{1/z}), \label{equ190}
\end{equation}
where the scaling function takes the form
\begin{equation}
   f_s(u) \sim  \left\{
   \begin{array}{lll}
     u^{2(\zeta-\zeta_s)}  & \quad &  \mbox{if ~ $u \gg 1 $} \\
     u^{2\zeta+1}  & \quad &  \mbox{if ~ $u \ll 1$}
   \end{array}\right. ,
   \label{equ200}
\end{equation}
and $\zeta_s$ is the {\it spectral} roughness exponent.

Since $\omega^2(t)$ describes the fluctuation in $x$ direction and
$M^{(2)}(t) - M(t)^2$ includes those in both $x$ and $y$ directions,
the dynamic exponent $z$ can be determined independently by
\begin{equation}
F(t) \sim \xi(t)/L_y \sim t^{1/z}/L_y. \label{equ210}
\end{equation}

\section{Monte Carlo Simulation}

We present the results with the random-single-spin-flip Monte Carlo algorithm
in the first two subsections, and those with different updating schemes in the third subsection.

\subsection{Uniform distribution of random fields}

\begin{figure}[p]
\setlength{\unitlength}{1.cm}
\begin{picture}(6,9)(0,0)
\includegraphics[width=7.cm]{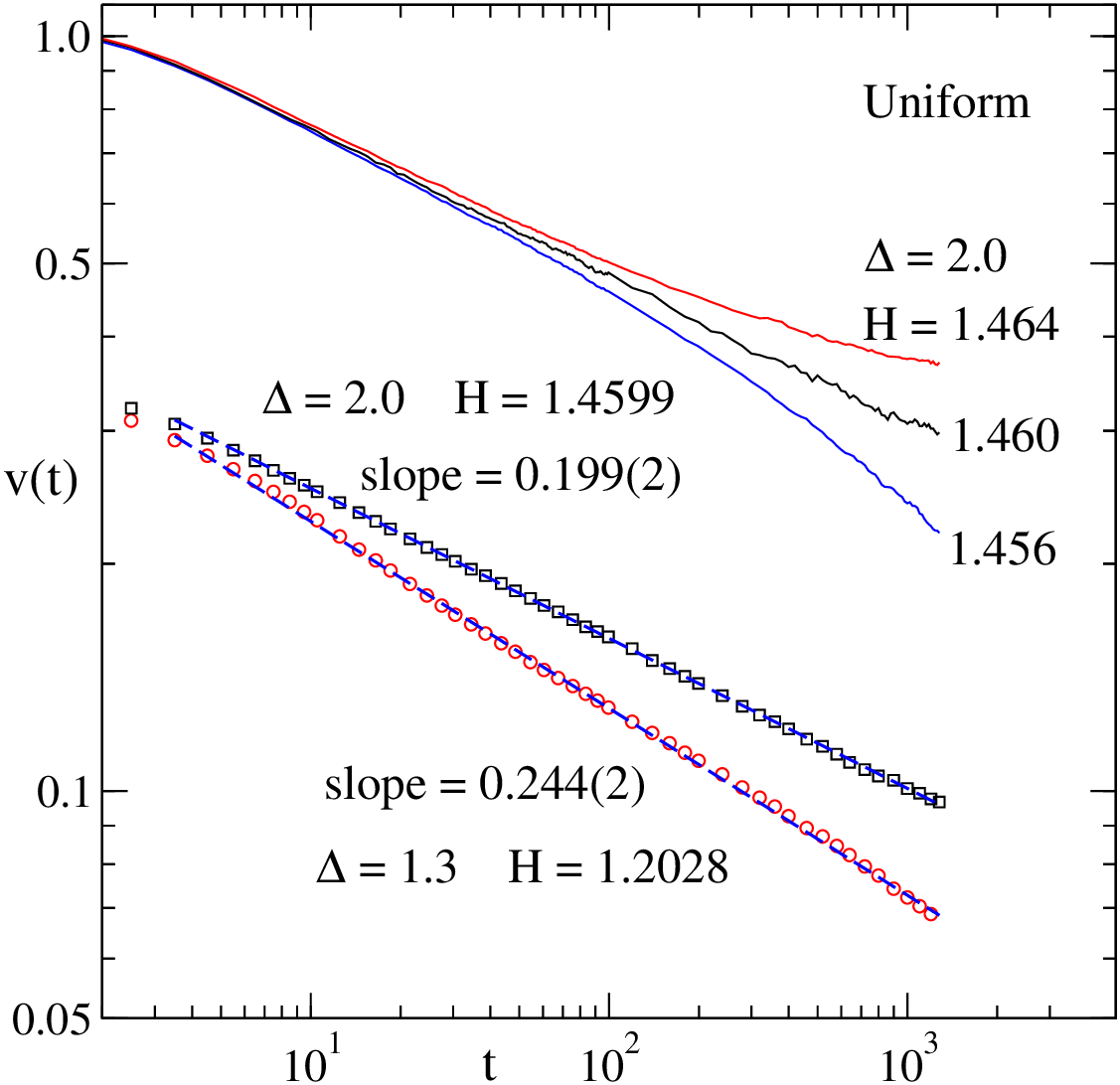}\quad\qquad
\includegraphics[width=7.cm]{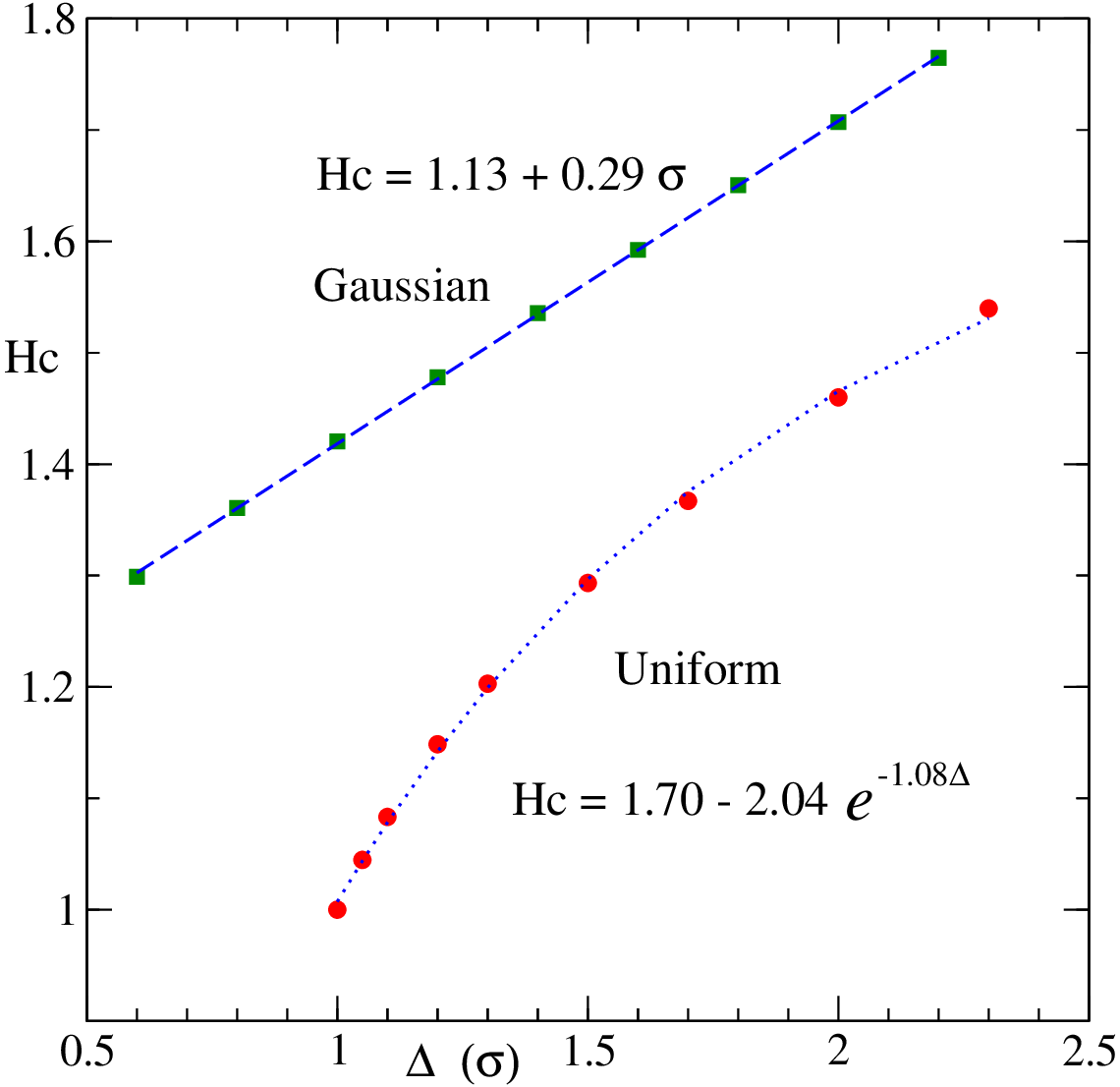}
\put(-14.8,-0.2){\footnotesize(a)} \put(-6.5,-0.2){\footnotesize(b)}
\end{picture}
\caption{(a) Interface velocity $v(t)$ is displayed for different
driving fields $H$ at $\Delta=2.0$ with solid lines for the uniform distribution of the random fields. For comparison,
$v(t)$ is also plotted for $\Delta=1.3$ and $2.0$ at $H_c=1.2028$
and $1.4599$ respectively with circles and squares. Dashed lines
show power-law fits. For clarity, the solid lines are shifted up.
(b) The transition field $H_c$ is
displayed, with $\Delta$ and $\sigma$ for the uniform and Gaussian
distributions of the random fields respectively. The dotted and dashed lines show exponential
and linear fits respectively. Errors are smaller than the
symbol sizes. } \label{vt}
\end{figure}

In Fig.~\ref{vt}(a), the interface velocity $v(t)$ is displayed for
different driving fields and different strengths of the random fields. For example, at $\Delta=2.0$, it drops
rapidly for smaller $H$, while approaches a constant for larger $H$.
By searching for the best power-law behavior, one locates the
transition fields $H_c=1.2028(2)$ and $1.4599(2)$ for $\Delta = 1.3$
and $2.0$, respectively. As shown in the figure, $v(t)$ for both
$\Delta = 1.3$ and $2.0$ at $H_c$ show almost perfect power-law
behaviors starting from rather early times. According to
Eq.~(\ref{equ130}), one measures the exponent $\beta /\nu
z=0.244(2)$ and $0.199(2)$ from the slopes of the curves for $\Delta
= 1.3$ and $2.0$, respectively.

The transition field $H_c$ varies significantly with the strength
of the random fields, as shown in
Fig.~\ref{vt}(b), and all the measurements of $H_c$ are summarized
in Table~\ref{t1}. The results are qualitatively
in agreement with the previous work~\cite{ji91}, but with much better accuracy.  For $\Delta
> 1$, the transition field $H_c < \Delta$, and the depinning transition is of second-order. The order
parameter $v(t)$ exhibits a nice power-law behavior, as shown in the
Fig.~\ref{vt}(a). For $\Delta \leq 1$, the velocity approaches a
constant for $H \geq \Delta$, while drops rapidly to zero once $H$
is slightly smaller than $\Delta$. This is a signal that the
depinning transition is of first-order for $\Delta \leq 1$, and the
transition field $H_c=\Delta$. Following Fig.~\ref{vt}(b), the
transition field $H_c$ increases with $\Delta$, and can be fitted to
an exponential function, i.e. $H_c=1.70-2.04 exp(-1.08\Delta)$.

\begin{figure}[p]
\setlength{\unitlength}{1.cm}
\begin{picture}(6,9)(0,0)
\includegraphics[width=7.cm]{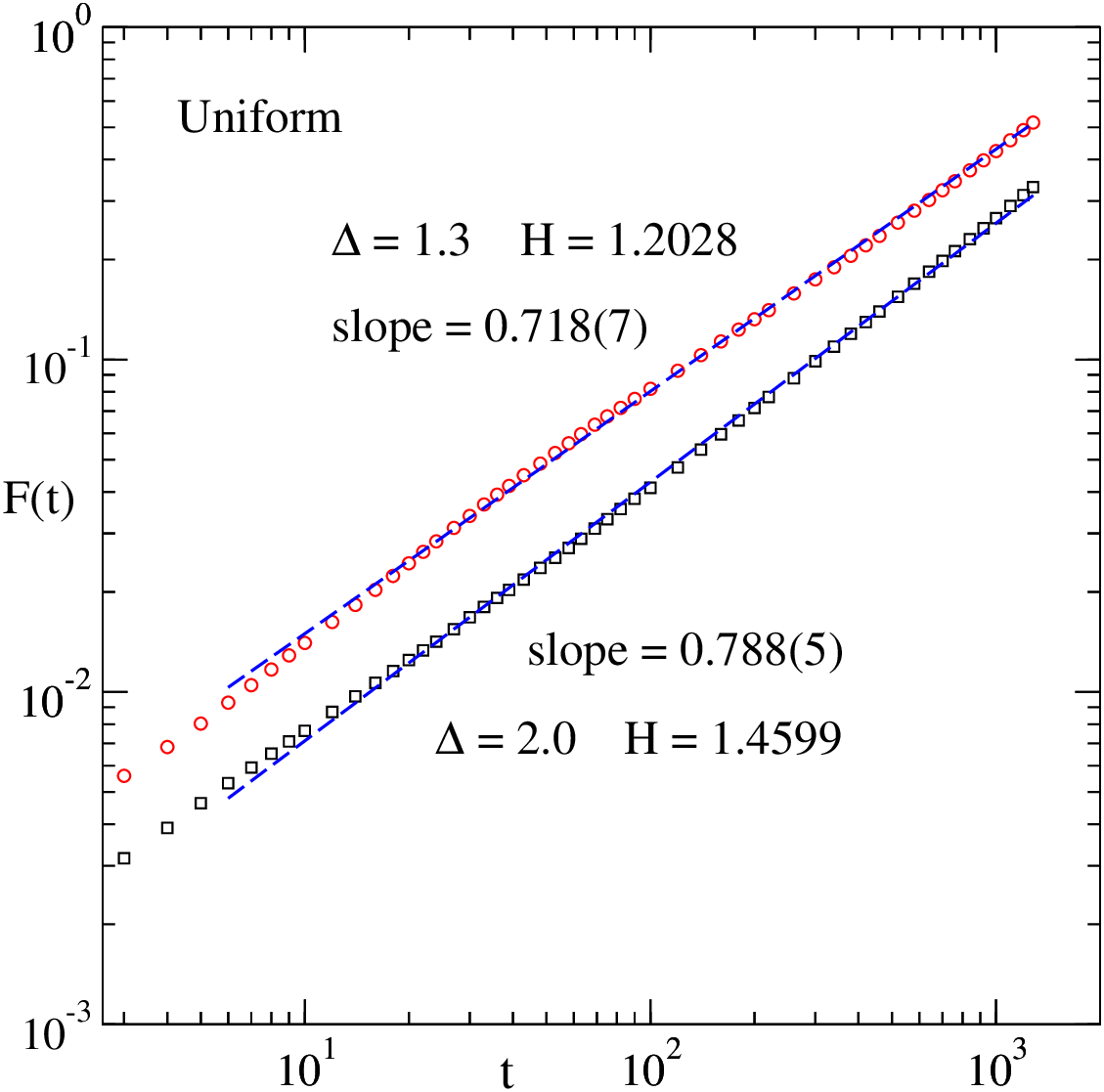}\quad\qquad
\includegraphics[width=7.2cm]{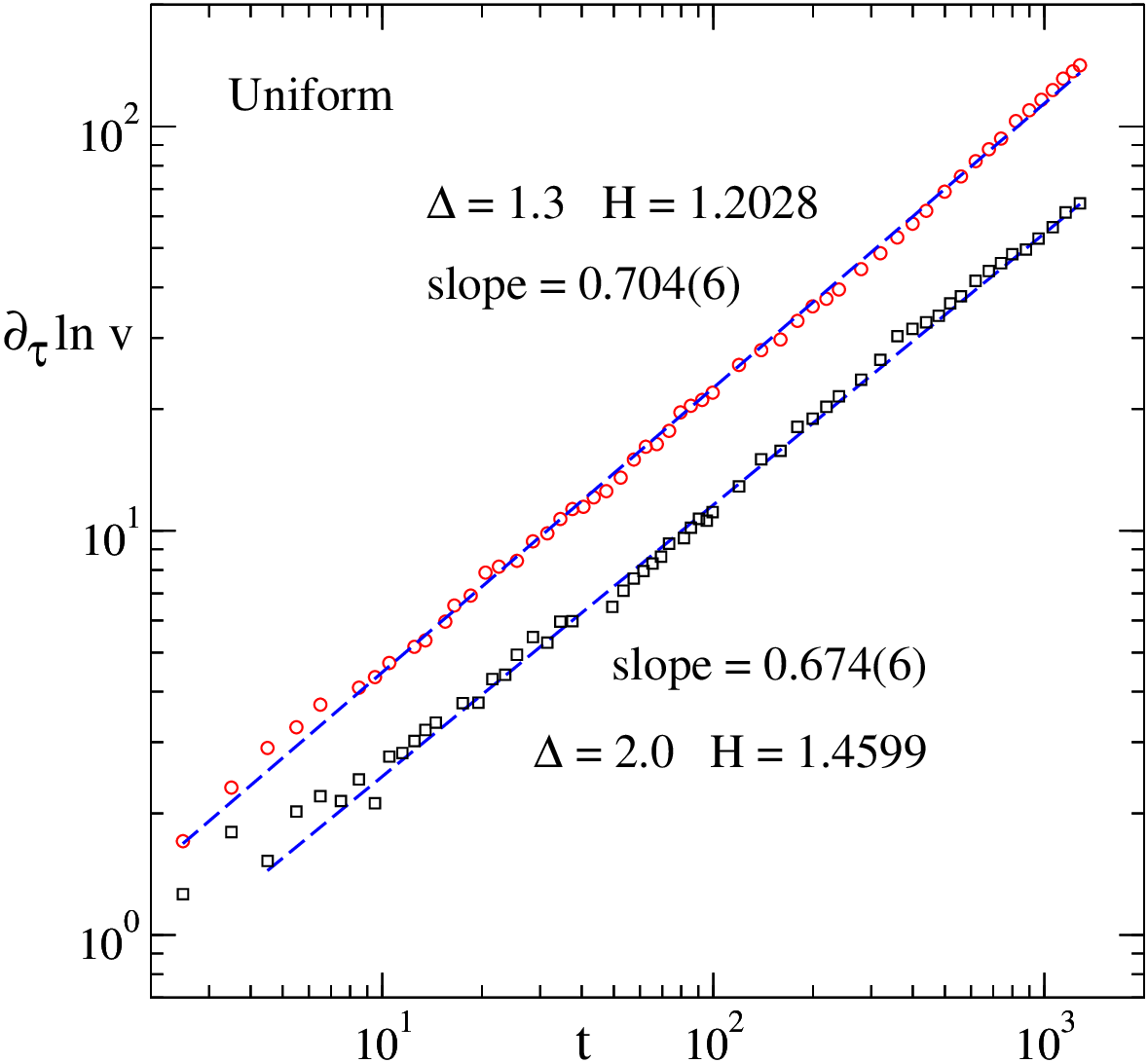}
\put(-14.7,-0.2){\footnotesize(a)} \put(-6.3,-0.2){\footnotesize(b)}
\end{picture}
\caption{(a) $F(t)$ is plotted for $\Delta = 1.3$ and $2.0$ at
$H_c$. For clarity, the curve for $\Delta = 1.3$ is shifted up. (b) The logarithmic
derivative of $v(t,\tau)$ is plotted for $\Delta = 1.3$ and $2.0$ at
$H_c$. In both (a) and (b), the results are for the uniform distribution of the random fields, and dashed lines show
power-law fits.} \label{Ft}
\end{figure}

In Fig.~\ref{Ft}(a), the dynamic observable $F(t)$ in
Eq.~(\ref{equ90}) is plotted at $H_c$. The power-law behavior is
detected. According to Eq.~(\ref{equ210}), $1/z=0.718(7)$ and
$0.788(5)$ are extracted respectively. To calculate the logarithmic
derivative $\partial _\tau \ln v(t,\tau) =\partial _\tau
v(t,\tau)/v(t,\tau)$, we interpolate $v(t,\tau)$ between different
$H$, e.g., in the interval $[1.195, 1.210]$ for $\Delta=1.3$, and
$[1.456, 1.464]$ for $\Delta=2.0$. In Fig.~\ref{Ft}(b), $\partial
_\tau \ln v(t,\tau)$ is plotted at $H_c$. A power-law behavior is
observed. Based on Eq.~(\ref{equ140}), $1/\nu z=0.704(6)$ and
$0.674(6)$ are derived from the slopes of the curves, respectively.

\begin{figure}[ht]
\setlength{\unitlength}{1.cm}
\begin{picture}(6,9)(0,0)
\includegraphics[width=7.cm]{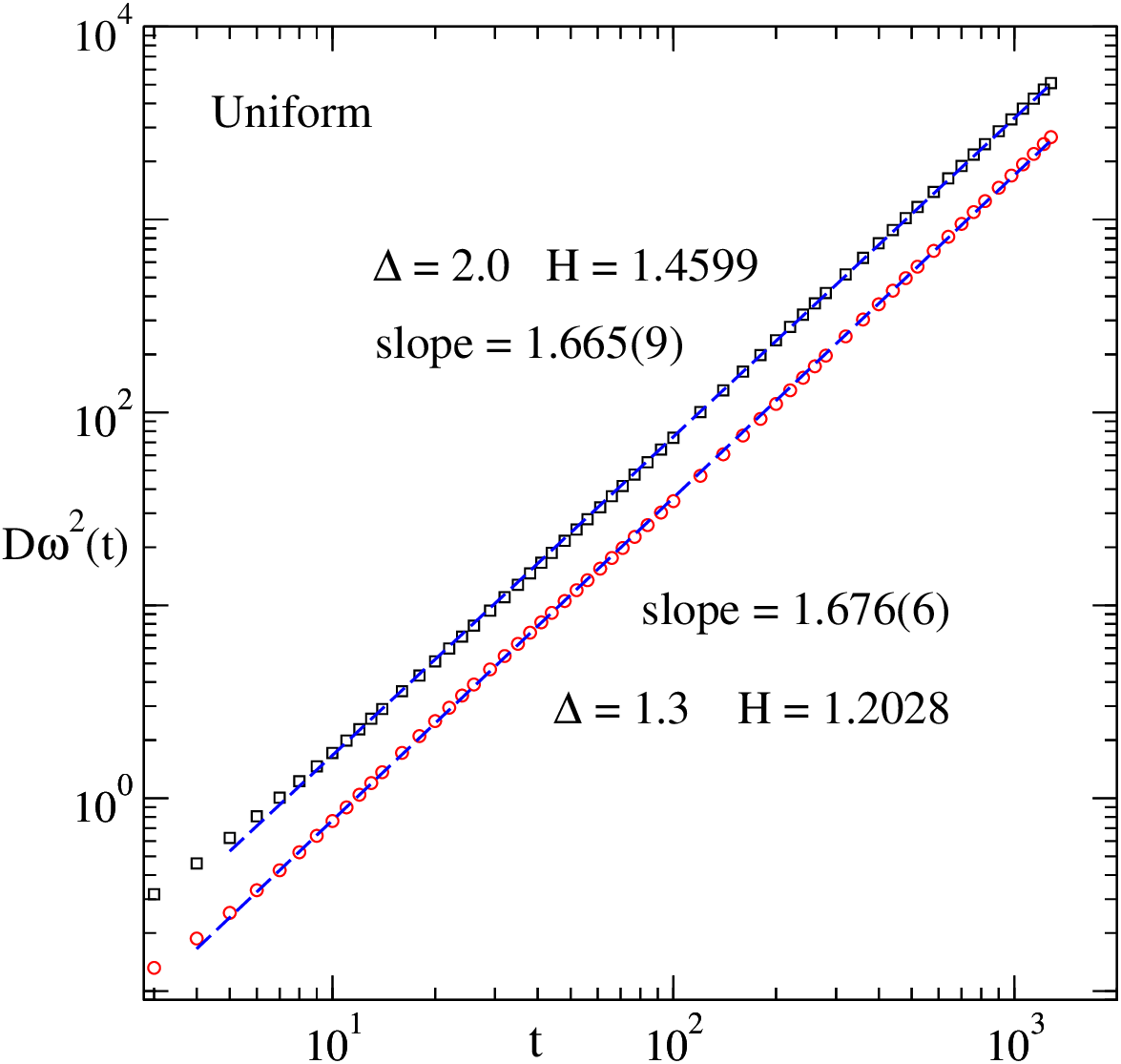}\quad\qquad
\includegraphics[width=7.2cm]{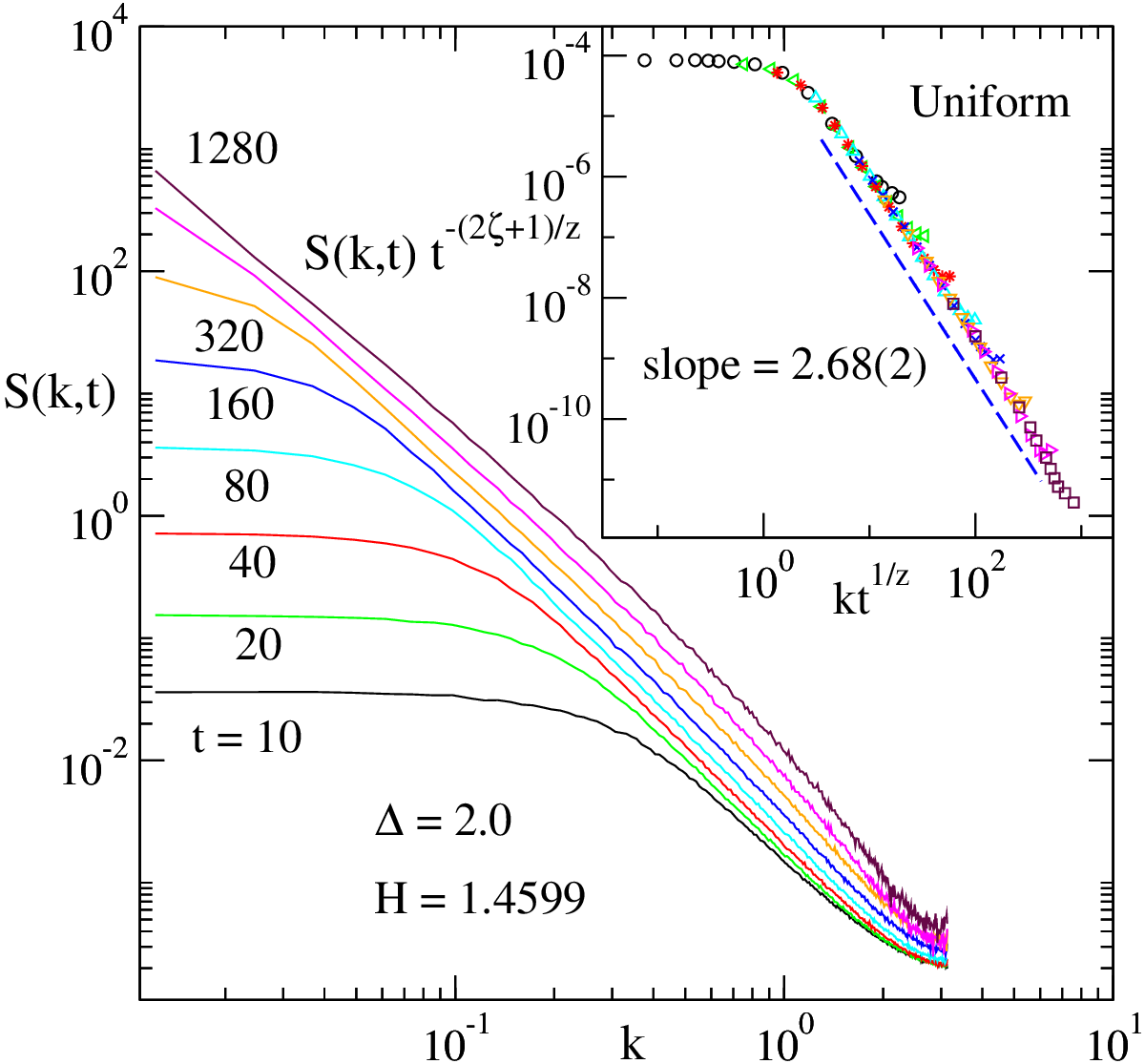}
\put(-14.5,-0.2){\footnotesize(a)} \put(-6.3,-0.2){\footnotesize(b)}
\end{picture}
\caption{(a) The pure roughness function $D\omega^2(t)$ is displayed
for $\Delta = 1.3$ and $2.0$ at $H_c$. (b) The structure factor
$S(k,t)$ is plotted for $\Delta=2.0$ at $H_c$. In the inset, data
collapse for different $t$ is shown. In both (a) and (b), the results are for the uniform distribution of the random fields, and dashed
lines show power-law fits. } \label{Dwt}
\end{figure}

In Fig.~\ref{Dwt}(a), the pure roughness function
$D\omega^2(t)=\omega^2(t) - \omega_b^2(t)$ is plotted for
$\Delta=1.3$ and $2.0$ at the transition field $H_c$. Here
$\omega^2_b(t)$ is the roughness function for $\Delta =0$. The
curves for both $\Delta = 1.3$ and $2.0$ show almost perfect
power-law behaviors from rather early times. According to
Eq.~(\ref{equ170}), we measure the exponent $2 \zeta / z=1.676(6)$
and $1.665(9)$ from the slopes of the curves respectively.

The structure factor $S(k,t)$ is displayed for $\Delta=2.0$ at
$H_c=1.4599$ in Fig.~\ref{Dwt}(b). According to Eq.~(\ref{equ200}),
data collapse for different $t$ is observed in the inset. The
exponents used for the data collapse are $\zeta=1.06$ and $z=1.27$
obtained from Fig.~\ref{Ft}(a) and \ref{Dwt}(a) respectively.
Obviously, for a small $u=kt^{1/z} \ll 1$, $u^{-(2\zeta+1)}f_s(u)$
approaches a constant. For a large $u \gg 1$,
$u^{-(2\zeta+1)}f_s(u)$ exhibits a power-law behavior. One extracts
the exponent $2\zeta_s +1 =2.68(2)$ from the slope of the curves,
i.e., $\zeta_s=0.84(1)$.

\begin{figure}[ht]
\setlength{\unitlength}{1.cm}
\begin{picture}(6,9)(0,0)
\includegraphics[width=7.cm]{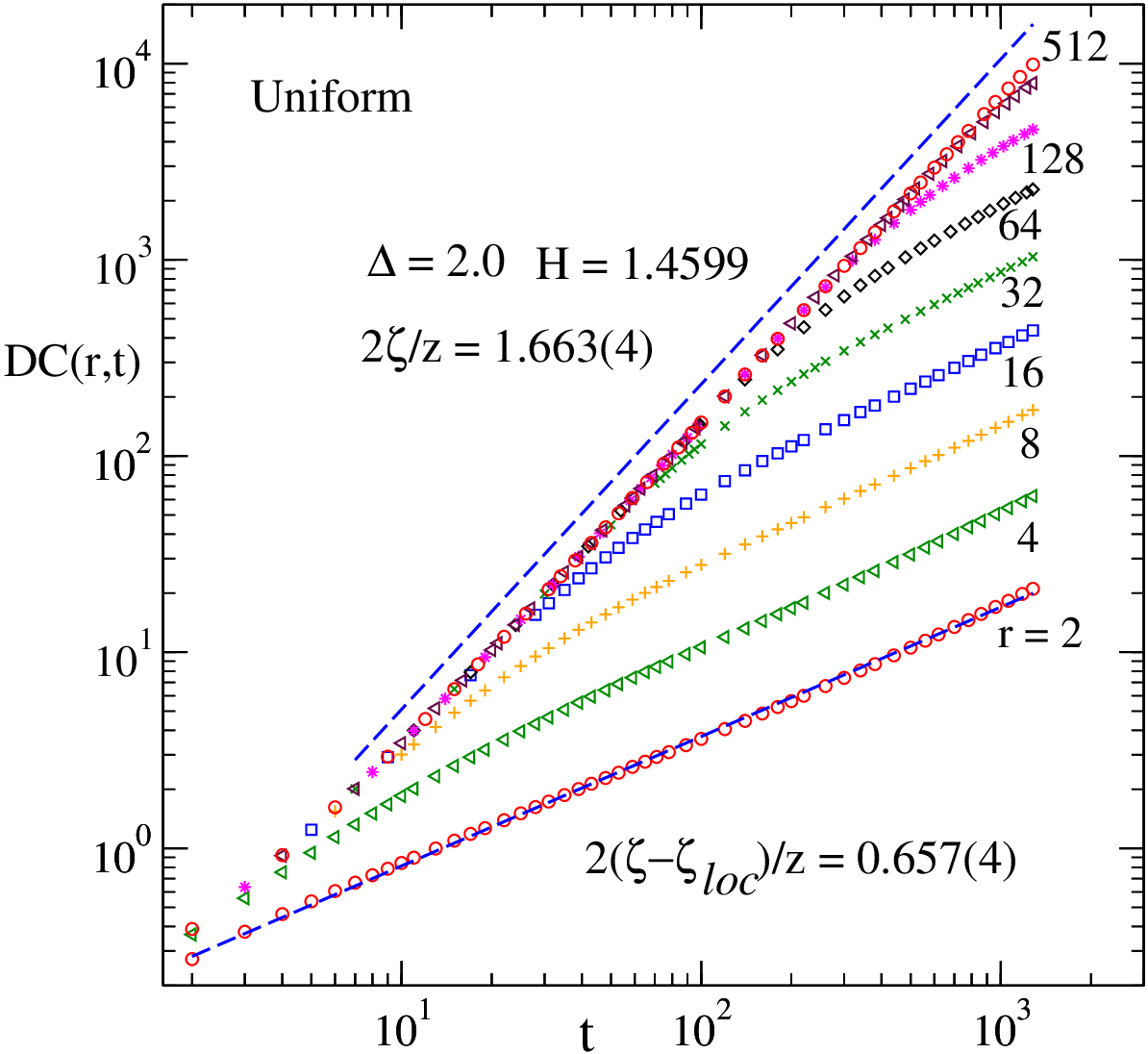}\quad\qquad
\includegraphics[width=7.1cm]{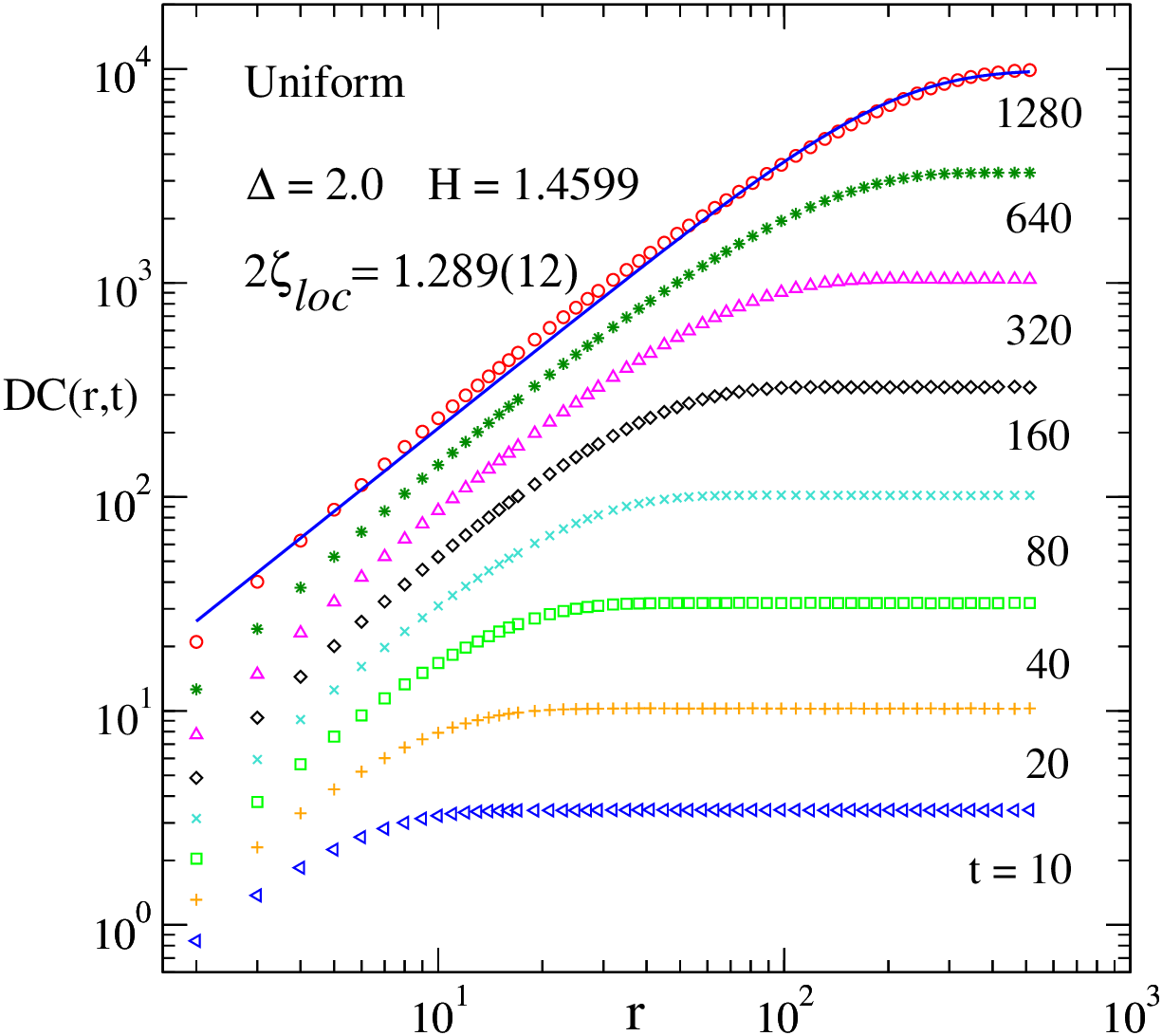}
\put(-14.4,-0.2){\footnotesize(a)} \put(-6.1,-0.2){\footnotesize(b)}
\end{picture}
\caption{(a) Time evolution of the pure height correlation
function $DC(r,t)$ is displayed for $\Delta=2.0$ at $H_c$.
Dashed lines show power-law fits. (b) $DC(r,t)$ is plotted as a
function of $r$ for $\Delta=2.0$ at $H_c$. The solid line
shows a power-law fit with a hyperbolic tangent correction. In both (a) and (b), the results are for the uniform distribution of the random fields.}
\label{dc}
\end{figure}

In Fig.~\ref{dc}(a), the pure height correlation function $DC(r,t)$
is displayed for $\Delta =2.0$ at $H_c=1.4599$. For a large $r\gg
\xi (t)$, e.g., $r=512$, one extracts the exponent $2\zeta /z
=1.663(4)$ by Eq.~(\ref{equ180}), consistent with that from
Fig.~\ref{Dwt}(a) within errors. For a small $r\ll \xi (t)$,
$DC(r,t)$ should be independent of $t$ for a normal interface with
$\zeta =\zeta_{loc}$, according to Eq.~(\ref{equ180}). But $DC(r,t)$
of $r=2$ clearly increases with time $t$. A power-law behavior is
observed from rather early times. According to Eq.~(\ref{equ180}),
the slope of the curve gives $2(\zeta-\zeta_{loc})/z =0.657(4)$.
From the measurements of $\zeta$ and $z$ from Fig.~\ref{Ft}(a) and
\ref{Dwt}(a), we calculate the local roughness exponent
$\zeta_{loc}=0.643(6)$.

The pure height correlation function $DC(r,t)$ is plotted as a
function of $r$ in Fig.~\ref{dc}(b). For a small $r\ll \xi
(t)$, there exists a power-law behavior, but this region is rather
limited. For a large $r$, there emerge strong corrections to scaling. We
introduce the form of corrections \cite{jos96},
\begin{equation}
DC(r) \sim [\tanh (r/c) ]^{2\zeta_{loc}}. \label{equ230}
\end{equation}
Here, $c$ is an adjustable parameter. The fitting to the curve extends
to a larger interval and yields $\zeta_{loc} =0.645(6)$. It is in
agreement with that from Fig.~\ref{dc}(a) within errors.

From the measurements of $\beta /\nu z$, $1/z$, $1/\nu z$, $2 \zeta
/ z$, $2\zeta_s +1$, $2(\zeta-\zeta_{loc})/z$ and $2 \zeta_{loc}$ in
Fig.~\ref{vt} to \ref{dc}, respectively. We calculate the dynamic
critical exponent $z$, the static critical exponents $\beta$ and
$\nu$, the roughness exponents $\zeta, \zeta_{loc}$ and $\zeta_s$.
All the measurements are summarized in Table~\ref{t1}, compared with
those of the QEW equation. In Fig.~\ref{univer}(a), the critical
exponents $z$, $\nu$, and $\zeta$ are displayed for different
strengths of the random fields. The dynamic critical exponent $z$
decreases with $\Delta$, and the difference between the DRFIM model
and QEW equation increases with $\Delta$, up to $20$ percent. The
critical exponent $\nu$ depends non-monotonously on $\Delta$, and
the difference between the DRFIM model and QEW equation reaches
maximally for $\Delta=1.3$ and $1.5$, over $20$ percent. In
Fig.~\ref{univer}(b), the critical exponent $\beta$ decreases
monotonously with $\Delta$ and reaches constant for $\Delta \geq
1.5$.

\begin{figure}[ht]
\setlength{\unitlength}{1.cm}
\begin{picture}(6,9)(0,0)
\includegraphics[width=7.cm]{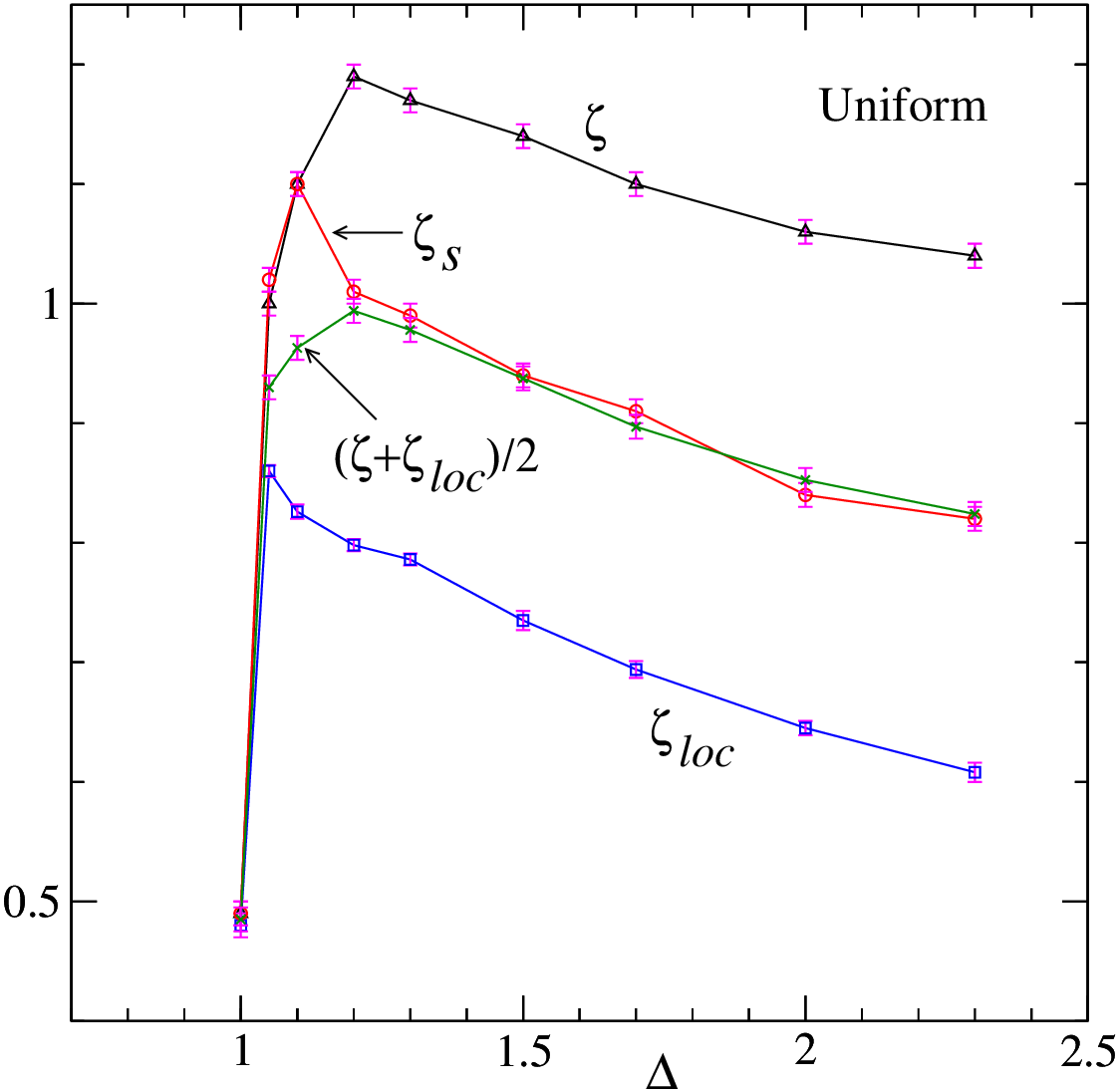}\quad\qquad
\includegraphics[width=7.cm]{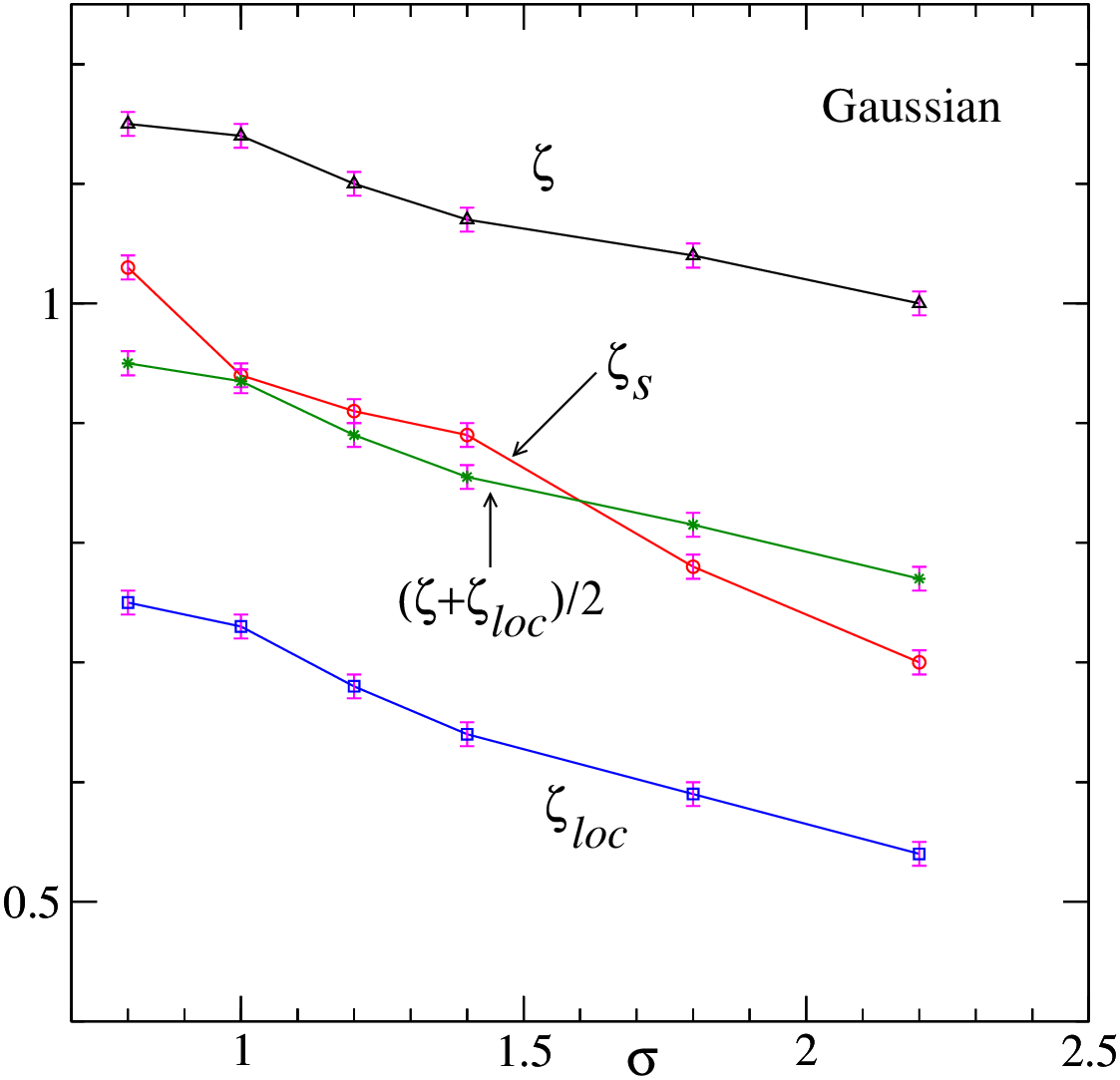}
\put(-14.8,-0.2){\footnotesize(a)} \put(-6.7,-0.2){\footnotesize(b)}
\end{picture}
\caption{ (a) The roughness exponents $\zeta, \zeta_{loc}, \zeta_s$
and $(\zeta+\zeta_{loc})/2$ are plotted for the uniform distribution of the
random fields. (b) The roughness exponents $\zeta,
\zeta_{loc}, \zeta_s$ and $(\zeta+\zeta_{loc})/2$ are displayed for the
Gaussian distribution of the random fields. In both (a) and
(b), full lines are just to guide the eyes. } \label{zeta}
\end{figure}

In Fig.~\ref{zeta}(a), the roughness exponents $\zeta, \zeta_{loc}$
and $\zeta_s$ are plotted for different strengths of the random fields.
According to Ref.~\cite{ram00}, the exponents $\zeta, \zeta_{loc}$
and $\zeta_s$ are not independent, and there are four different
subclasses, namely, Family-Vicsek, super-rough, intrinsic and
faceted. Following Fig.~\ref{zeta}(a) and Table~\ref{t1}, $\zeta=
\zeta_{loc} =\zeta_s$ for $\Delta \leq 1$, and it belongs to the
{\it Family-Vicsek} universality class. In the crossover regime from
the first-order transition to the second-order one, e.g., for
$\Delta=1.05$ and $1.1$, $\zeta=\zeta_s$, the DRFIM model looks
somewhat similar to the QEW equation, but there exists still a major
difference, i.e., $\zeta_{loc}<1$ for the DRFIM model, while
$\zeta_{loc} \approx 1$ for the QEW equation. For $\Delta \geq 1.2$,
$\zeta \neq \zeta_{loc}\neq \zeta_s$ and $\zeta_{loc} \neq 1$, and
it indicates that the DRFIM model belongs to the new universality
class suggested in Ref.~\cite{che10}. Additionally, one
interestingly observes that the exponent $\zeta_s$ is in agreement
with $(\zeta+\zeta_{loc})/2$ within errors. The possible mechanism
remains to be understood.

For $\Delta>1$, the local roughness exponent $\zeta_{loc}$ decreases
with $\Delta$, and the difference between the DRFIM model and QEW
equation increases with $\Delta$, up to over $35$ percent. In our
opinions, it is mainly the overhangs and islands that induce this
difference.

\subsection{Gaussian distribution of random fields}

\begin{figure}[ht]
\setlength{\unitlength}{1.cm}
\begin{picture}(6,9)(0,0)
\includegraphics[width=7.cm]{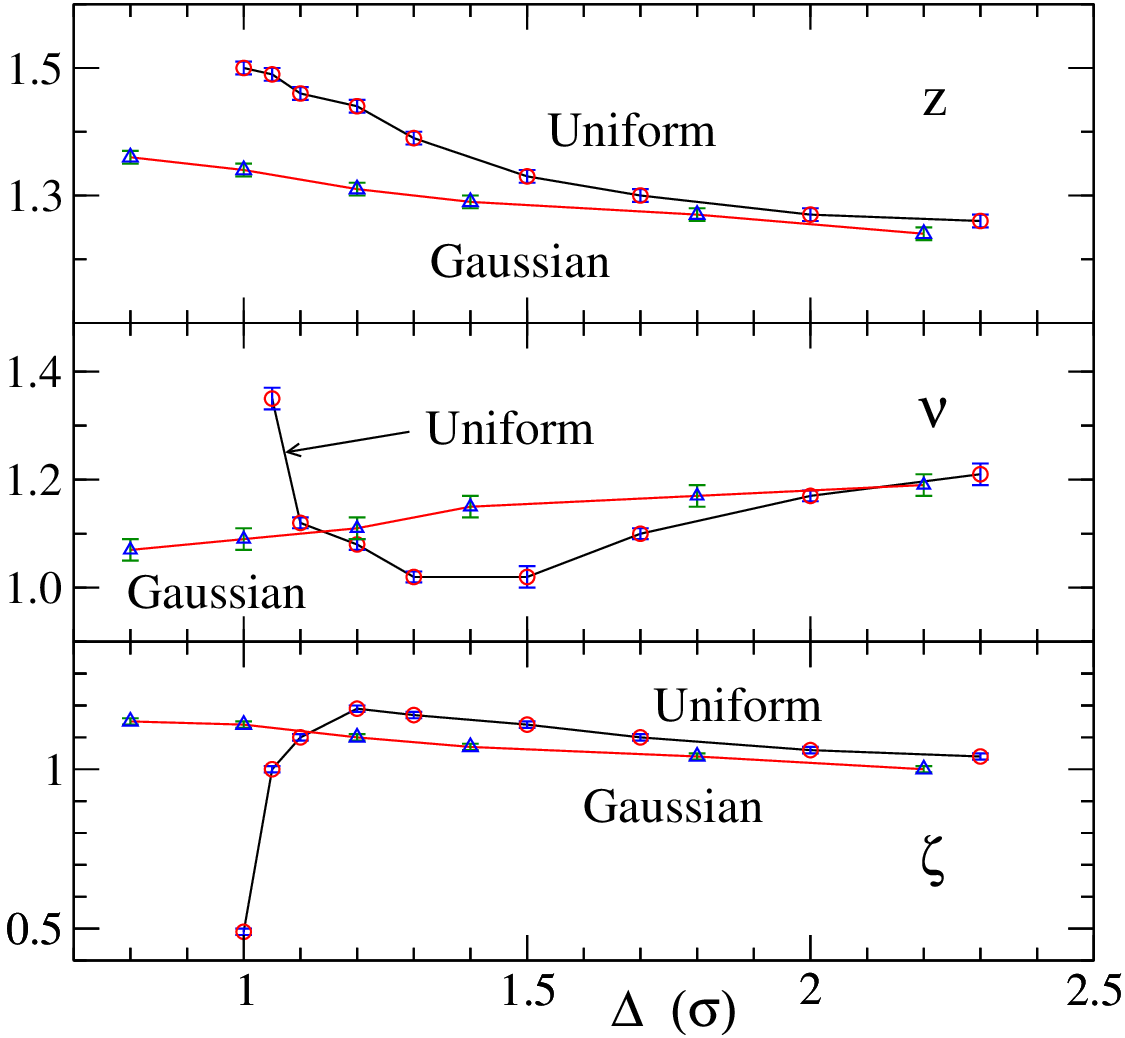}\quad\qquad
\includegraphics[width=7.cm]{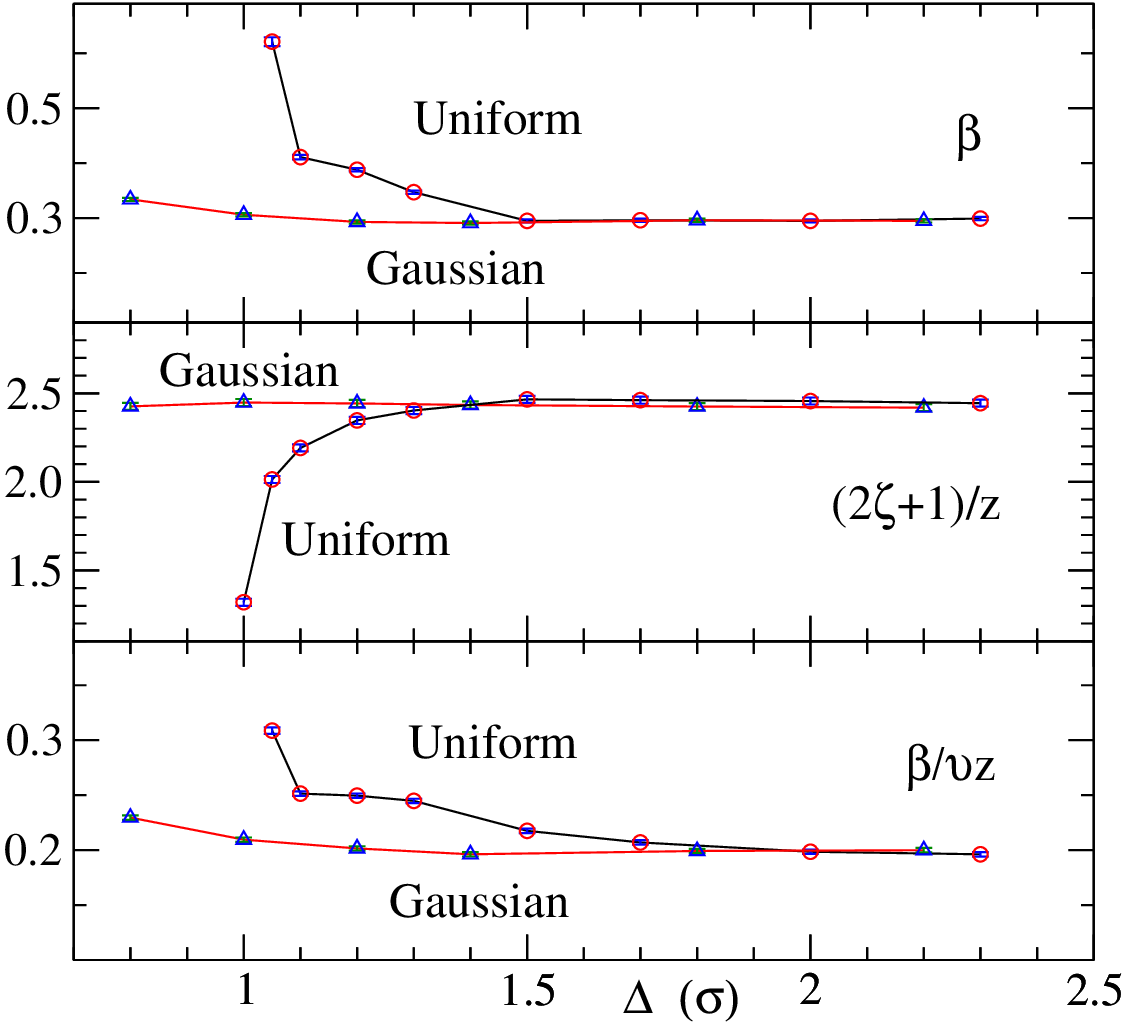}
\put(-14.8,-0.2){\footnotesize(a)} \put(-6.7,-0.2){\footnotesize(b)}
\end{picture}
\caption{In (a) and (b), variation of the critical exponents $z$, $\nu$,
$\zeta$, $\beta$, $(2\zeta+1)/z$ and $\beta/\nu z$ with $\Delta$ and $\sigma$ respectively for the uniform and
Gaussian distributions of the random fields is displayed. } \label{univer}
\end{figure}

As shown in Fig.~\ref{vt}(b), the transition field $H_c$ varies linearly with $\sigma$
for the Gaussian distribution of the random fields, and all the measurements of $H_c$ are summarized
in Table~\ref{t2}. The results
are in agreement with those in Ref.~\cite{ji91}, but with higher accuracy.
Here the depinning transition is always of second-order. The crossover
from the second-order phase transition to the first-order one is not
present, because the Gaussian distribution is
unbounded.
All the measurements of the critical exponents are also summarized in Table~\ref{t2}.
In Fig.~\ref{univer}(a), variation of the critical exponents $z$,
$\nu$, and $\zeta$ with $\sigma$ for the Gaussian distribution of the random
fields is displayed. The dynamic critical exponent $z$ decreases
monotonously with $\sigma$, similar to that of the uniform
distribution of the random fields. However, the static critical exponent $\nu$
increases monotonously with $\sigma$, different from that of the uniform distribution of the random fields.
In our opinions, the non-monotonous behavior of the critical exponents for the uniform distribution of the random fields
is induced by the crossover
from the second-order transition to the first-order one.
In Fig.~\ref{zeta}(b), the monotonous decrease of the roughness
exponents $\zeta, \zeta_{loc}$ and $\zeta_s$ with $\sigma$ is displayed for the Gaussian distribution of the random fields.
The results indicate $\zeta
\neq \zeta_{loc}\neq \zeta_s$ and $\zeta_{loc} \neq 1$, the
same as that of the uniform distribution of the random fields.

As shown in Figs.~\ref{univer} and \ref{zeta}, all the critical exponents
$z$, $\nu$, $\beta$, $\zeta$, $\zeta_{loc}$ and $\zeta_s$ vary in principle with
the strength of the random fields, and the dependence of the critical exponents
on the strength of the random fields is also different for the uniform and Gaussian distributions.
At the limit of the strong random fields, the critical exponents
may tend to the same for the uniform and Gaussian distributions, but it is less clear
for the roughness exponents $\zeta_{loc}$ and $\zeta_s$.
In other words, the strong universality is violated in the depinning transition.
Both the crossover from the second-order transition to the first-order one in the uniform distribution of the random fields,
and the dynamic effect of the overhangs and islands lead to this violation of the universality.
It is pointed out in Ref. \cite{zho10} that overhangs and islands play important roles in the domain-wall motion.
As the strength of the random fields changes, the overhangs and islands alter their critical behavior
and modify the critical exponents at the depinning transition.

Although the strong universality is violated, we observe that the static exponent $\beta$ looks somewhat 'universal'.
For the Gaussian distribution of the random fields,
$\beta$ is almost independent of the strength of the random fields.
For the uniform distribution of the random fields, $\beta$ also reaches a constant value for $\Delta\geq 1.5$
which coincides with that of the Gaussian distribution.
Are there other combinations of the critical exponents which show a 'universal' behavior?
In Fig.~\ref{univer}(b), variation of the exponents $(2\zeta+1)/z$ and $\beta/\nu z$
with the strength of the random fields is displayed.
In particular, the exponent $(2\zeta+1)/z$ looks 'universal', similar to the exponent $\beta$.
According to
Eqs.~(\ref{equ170}) and (\ref{equ210}), $M^{(2)}(t) - M(t)^2 \sim
t^{(2\zeta+1)/2}$. In other words, the scaling exponent of the $planar$
susceptibility is 'universal'. Since $M(t)$ is equivalent to the average
height function, $M^{(2)}(t) - M(t)^2$ is nothing but
the fluctuation of the domain wall.

\subsection{Updating schemes}

To further understand the universality at the depinning transition,
we examine different update schemes in the Monte Carlo simulations, e.g.,
the sequential sweep and parallel sweep.
For the $sequential$ sweep,
we select a spin sequentially by the row perpendicular to the domain wall, and flip it if the total energy decreases after flipping.
For the $parallel$ sweep, we divide the lattice into two 'checkerboard' sublattices.
The spins in each sublattice do not directly interact each other.
Therefore we may update all the spins in each sublattice simultaneously.
Different updating schemes in the Monte Carlo algorithm do not modify the transition field and critical exponents,
and all the measurements are summarized in Table~\ref{t3}.
The small deviation in the transition field $H_c$ and the exponent $\beta$
should be induced by corrections to scaling in the early times.

Because of the presence of the random fields, the magnetic system
has many local minima separated by sizeable energy barriers.
During the dynamic evolution, the system may be trapped in a local minimum without finding the global one,
as a consequence, remains far from equilibrium.
The lifetime of a metastable state may be considered as infinite at zero-temperature.
This type of behavior is well modeled by the single-spin-flip Monte Carlo dynamics \cite{viv05}.
In ordered to verify the reliability of the single-spin-flip Monte Carlo dynamics,
the two-spin-flip algorithm has been also performed.
The power-law behaviors in Fig~\ref{vt} to \ref{dc} are
observed. All the measurements of the depinning transition field and
critical exponents are summarized in Table~\ref{t3}.
The depinning transition field $H_c$ is changed,
but all the exponents are in agreement within errors with those of the single-spin-flip dynamics \cite{viv05}.

\section{Conclusion}

With Monte Carlo methods, we have simulated the dynamic relaxation of
a domain wall in the two-dimensional DRFIM model.
Based on the short-time dynamic scaling forms, the transition field
and all critical exponents at the depinning transition are accurately
determined. Since the measurements are carried out at the beginning
of the time evolution, the dynamic approach
does not suffer from critical slowing down.

The results are first presented for the uniform
distribution of the random fields with the single-spin-flip Monte Carlo dynamics.

i) For $\Delta \leq 1$, the depinning transition is of first-order, and the
transition field $H_c=\Delta$. The exponents $\beta$ and $\nu$ do
not exist, but the growth dynamics of the domain wall is still
meaningful. Since $\zeta= \zeta_{loc} =\zeta_s$, it belongs to the
{\it Family-Vicsek} universality class.

ii) For $\Delta > 1$, the depinning transition is of second-order, and
the transition field $H_c < \Delta$. As shown in
Fig.~\ref{univer}(a) and Fig.~\ref{zeta}(a), critical exponents vary
significantly with $\Delta$. Except for the dynamic exponent $z$,
all other critical exponents indicate the crossover from the
first-order transition to the second-order one. The exponent $\nu$
and roughness exponents exhibit non-monotonous dependence on
$\Delta$.

iii) For $\Delta > 1$ and $\Delta
\rightarrow 1$, e.g., at $\Delta=1.05$ and $1.1$, the DRFIM model
looks somewhat similar to the QEW equation, but $\zeta_{loc} \neq 1$.
If $\zeta_{loc}=1$, it would be in the so-called {\it super-rough} universality class as the QEW equation is.
For $\Delta \geq 1.2$, one observes clearly $\zeta \neq \zeta_{loc}\neq \zeta_s$,
and $\zeta_{loc}<1$. It
indicates that the DRFIM model belongs to the new dynamic universality class suggested in Ref.~\cite{che10}.

For comparison, simulations have been also performed for the Gaussian
distribution of the random fields. The crossover from the second-order
transition to the first-order one is not present in this case.
Generally speaking, the strong universality is violated in the depinning transition, and
the critical exponents vary significantly with
the strength and form of the random fields.
In our opinion, the difference between the uniform and Gaussian distributions of the random fields
is mainly induced by the crossover from the second-order
transition to the first-order one in the former case, and the dependence of the critical exponents
on the strength of the random fields comes also from the overhangs and islands.
Finally, our simulations confirm that the critical
exponents are independent of the updating schemes.

\ack  This work was supported in part by NNSF of China under Grant
Nos. 10875102 and 11075137, and Research Projects of Education Department of
Zhejiang No. Y200803742.

\section*{References}


\begin{thebibliography}{10}

\bibitem{nat01}
{Nattermann T, Pokrovsky V and Vinokur V M} 2001
\newblock {\em Phys. Rev. Lett.} {\bf 87} 197005

\bibitem{gla03}
{Glatz A, Nattermann T and Pokrovsky V} 2003
\newblock {\em Phys. Rev. Lett.} {\bf 90} 047201

\bibitem{ros03}
{Rosso A, Hartmann A K and Krauth W} 2003
\newblock {\em Phys. Rev.} E {\bf 67} 021602

\bibitem{due05}
{Duemmer O and Krauth W} 2005
\newblock {\em Phys. Rev.} E {\bf 71} 061601

\bibitem{kol06}
{Kolton A B, Rosso A, Albano E V and Giamarchi T} 2006
\newblock {\em Phys. Rev.} B {\bf 74} 140201(R)

\bibitem{kol06a}
{Kolton A B, Rosso A, Giamarchi T and Krauth W} 2006
\newblock {\em Phys. Rev. Lett.} {\bf 97} 057001

\bibitem{kle07a}
{Kleemann W} 2007
\newblock {\em Annu. Rev. Mater. Res.} {\bf 37} 415

\bibitem{bus08}
{Bustingorry S, Kolton A B and Giamarchi T} 2008
\newblock {\em Europhys. Lett.} {\bf 81} 26005

\bibitem{now98}
{Nowak U and Usadel K D} 1998
\newblock {\em Europhys. Lett.} {\bf 44} 634

\bibitem{rot99}
{Roters L, Hucht A, L\"ubeck S, Nowak U and Usadel K D} 1999
\newblock {\em Phys. Rev.} E {\bf 60} 5202

\bibitem{rot01}
{Roters L, L\"ubeck S and Usadel K D} 2001
\newblock {\em Phys. Rev.} E {\bf 63} 026113

\bibitem{zho09}
{Zhou N J, Zheng B and He Y Y} 2009
\newblock {\em Phys. Rev.} B {\bf 80} 134425

\bibitem{zho10}
{Zhou N J and Zheng B} 2010
\newblock {\em Phys. Rev.} E {\bf 82} 031139

\bibitem{met07}
{Metaxas P J, Jamet J P and Mougin A {\em et al}} 2007
\newblock {\em Phys. Rev. Lett.} {\bf 99} 217208

\bibitem{lem98}
{Lemerle S, Ferr\'{e} J and Chappert C {\em et al}} 1998
\newblock {\em Phys. Rev. Lett.} {\bf 80} 849

\bibitem{jos98}
{Jost M, Heimel J and Kleinefeld T} 1998
\newblock {\em Phys. Rev.} B {\bf 57} 5316

\bibitem{ros01}
{Rosso A and Krauth W} 2001
\newblock {\em Phys. Rev. Lett.} {\bf 87} 187002

\bibitem{zhe98}
{Zheng B} 1998
\newblock {\em Int. J. Mod. Phys.} B {\bf 12} 1419
\newblock (review article)

\bibitem{luo98}
{Luo H J, Sch\"ulke L and Zheng B} 1998
\newblock {\em Phys. Rev. Lett.} {\bf 81} 180

\bibitem{rod07}
{Rodr\'{i}guez-Rodr\'{i}guez G and P\'{e}rez-Junquera A {\em et al}}
2007
\newblock {\em J. Phys. D: Appl. Phys.} {\bf 40} 3051

\bibitem{kol09}
{Kolton A B, Schehr G and Le Doussal P} 2009
\newblock {\em Phys. Rev. Lett.} {\bf 103} 160602

\bibitem{ram00}
{Ramasco J J, L\'{o}pez J M and Rodr\'{\i}guez M A} 2000
\newblock {\em Phys. Rev. Lett.} {\bf 84} 2199

\bibitem{che10}
{Chen Y J, Nagamine Y, Yamaguchi T and Yoshikawa K} 2010
\newblock {\em Phys. Rev.} E {\bf 82} 021604

\bibitem{ji91}
{Ji H and Robbins M O} 1991
\newblock{\em Phys. Rev.} A {\bf 44} 2538

\bibitem{koi10}
{Koiller B and Robbins M O} 2010
\newblock{\em Phys. Rev.} B {\bf 82} 064202

\bibitem{dro98}
{Drossel B and Dahmen K} 1998
\newblock{\em Eur. Phys. J.} B {\bf 3} 485

\bibitem{viv05}
{Vives E, Rosinberg M L and Tarjus G} 2005
\newblock{\em Phys. Rev.} B {\bf 71} 134424

\bibitem{jan89}
{Janssen H K, Schaub B and Schmittmann B} 1989
\newblock {\em Z. Phys.} B {\bf 73} 539

\bibitem{jos96}
{Jost M and Usadel K D} 1996
\newblock {\em Phys. Rev.} B {\bf 54} 9314

\bibitem{zho08}
{Zhou N J and Zheng B} 2008
\newblock {\em Phys. Rev.} E {\bf 77} 051104

\bibitem{bak08}
{Bak\'o B, Weygand D, Samaras M, Hoffelner W and Zaiser M} 2008
\newblock {\em Phys. Rev.} B {\bf 78} 144104

\end{thebibliography}

\begin{table}[h]
\begin{tabular}[t]{ c | c c c | c | c| c | c }
\hline \hline
                 &     &    $v(t)$      &         & $F(t)$,$C(r,t)$  & $\omega^2(t)$,$C(r,t)$ & $C(r,t)$      & $S(k,t)$ \\
\hline

$\Delta$         &$H_c$      &$\beta$   & $\nu$   & $z$     & $\zeta$  & $\zeta_{loc}$  & $\zeta_s$   \\
\hline
$\leq 1.0$       &$\Delta$   &          &         & 1.50(1) & 0.49(1)  & 0.48(2)  & 0.49(1)  \\
1.05             & 1.0447(1) & 0.621(8) & 1.35(3) & 1.49(1) & 1.00(1)  & 0.86(1)  & 1.02(2)  \\
1.1              & 1.0833(2) & 0.411(4) & 1.12(2) & 1.46(1) & 1.10(1)  & 0.83(1)  & 1.10(1)  \\
1.2              & 1.1485(2) & 0.388(3) & 1.08(2) & 1.44(1) & 1.19(1)  & 0.80(1)  & 1.01(1)  \\
1.3              & 1.2028(2) & 0.347(3) & 1.02(2) & 1.39(1) & 1.17(1)  & 0.78(1)  & 0.99(1)  \\
1.5              & 1.2933(2) & 0.295(3) & 1.02(3) & 1.33(1) & 1.14(1)  & 0.74(1)  & 0.94(1)  \\
1.7              & 1.3670(2) & 0.296(3) & 1.10(2) & 1.30(1) & 1.10(1)  & 0.69(1)  & 0.91(1)  \\
2.0              & 1.4599(2) & 0.295(3) & 1.17(2) & 1.27(1) & 1.06(1)  & 0.65(1)  & 0.84(1)  \\
2.3              & 1.5398(2) & 0.299(3) & 1.21(2) & 1.26(1) & 1.04(1)  & 0.61(1)  & 0.82(1)  \\
\hline
QEW              &           & 0.33(2)  & 1.33(4) & 1.50(3) & 1.25(1)  & 0.98(6)  & 1.25(1)  \\
\hline \hline
\end{tabular}
\caption{The depinning transition field and critical exponents
obtained for different strengths of the random fields with the uniform
distribution are compared with those of the QEW equation in
Refs.~\cite{due05,kol06,bus08}. } \label{t1}
\end{table}

\begin{table}[h]
\begin{tabular}[t]{ c | c c c | c | c | c | c  }
\hline \hline
                 &     &    $v(t)$      &         & $F(t)$,$C(r,t)$  & $\omega^2(t)$,$C(r,t)$ & $C(r,t)$    & $S(k,t)$ \\
\hline

$\sigma$         &$H_c$      &$\beta$   & $\nu$   & $z$     & $\zeta$  & $\zeta_{loc}$  & $\zeta_s$   \\
\hline
0.8              & 1.3607(3) & 0.334(3) & 1.07(2) & 1.36(1) & 1.15(1)  & 0.75(1)  & 1.03(1)  \\
1.0              & 1.4205(3) & 0.306(3) & 1.09(2) & 1.34(1) & 1.14(1)  & 0.73(1)  & 0.94(1)  \\
1.2              & 1.4779(3) & 0.293(3) & 1.11(2) & 1.31(1) & 1.10(1)  & 0.68(1)  & 0.91(1)  \\
1.4              & 1.5355(3) & 0.291(3) & 1.15(2) & 1.29(1) & 1.07(1)  & 0.64(1)  & 0.89(1)  \\
1.8              & 1.6505(3) & 0.296(3) & 1.17(2) & 1.27(1) & 1.04(1)  & 0.59(1)  & 0.78(1)  \\
2.2              & 1.7647(3) & 0.295(3) & 1.19(2) & 1.24(1) & 1.00(1)  & 0.54(1)  & 0.70(1)  \\
\hline \hline
\end{tabular}
\caption{The depinning transition field and critical exponents for the Gaussian distribution of the random fields. }
\label{t2}
\end{table}

\begin{table}[h]
\begin{tabular}[t]{ c | c | c c c | c | c| c | c }
\hline \hline

    &  sweep    &$H_c$      &$\beta$   & $\nu$   & $z$     & $\zeta$  & $\zeta_{loc}$  & $\zeta_s$      \\

\hline
    & random           & 1.2028(2) & 0.347(3) & 1.02(2) & 1.39(1) & 1.17(1)  & 0.78(1)  & 0.99(1)  \\
SSF & sequential       & 1.2034(4) & 0.345(6) & 1.02(2) & 1.40(1) & 1.17(1)  & 0.77(1)  & 0.99(1)  \\
    & parallel         & 1.2033(4) & 0.338(6) & 1.00(2) & 1.40(1) & 1.17(1)  & 0.77(1)  & 0.99(1)  \\
\hline
TSF & random           & 0.7545(3) & 0.340(6) & 1.01(2) & 1.40(2) & 1.19(2)  & 0.78(2)  & 1.01(2)  \\
\hline \hline
\end{tabular}
\caption{The depinning transition field and critical exponents for the uniform distribution of the
random fields at $\Delta=1.3$ with different update schemes. SSF: the single-spin-flip; TSF: the two-spin-flip. } \label{t3}
\end{table}

\end{document}